\documentclass[twocolumn]{aastex631}

\usepackage{comment}
\usepackage{ulem}
\usepackage{tikz}
\usepackage{color}
\usepackage{graphicx}
\usepackage{amsmath}
\usepackage{amsfonts}
\usepackage{booktabs}
\usepackage{multirow}
\usepackage{array}
\usepackage{tablefootnote}
\usepackage{threeparttable}

\usepackage[hang,flushmargin]{footmisc}
\usepackage[T1]{fontenc}
\usepackage[utf8]{inputenc}

\usepackage{listings}

\definecolor{dkgreen}{rgb}{0,0.6,0}
\definecolor{gray}{rgb}{0.5,0.5,0.5}
\definecolor{mauve}{rgb}{0.58,0,0.82}

\newcommand{\beq}{\begin{equation}}
\newcommand{\eeq}{\end{equation}}
\newcommand{\bea}{\begin{eqnarray}}
\newcommand{\eea}{\end{eqnarray}}

\newcommand{\gv}[1]{\ensuremath{\mbox{\boldmath$ #1 $}}} 
\newcommand{\trm}[1]{\textrm{#1}}

\newcommand{\non}{\nonumber \\}

\newcommand{\grad}[1]{\gv{\nabla} #1}

\newcommand{\T}{\ensuremath{\tau}}
\newcommand{\aveT}{\ensuremath{\langle\T\rangle}}
\newcommand{\Q}{\ensuremath{Q_\star^\prime}}
\newcommand{\Porb}{\ensuremath{P_{\rm orb}}}
\newcommand{\Mj}{\ensuremath{M_{\rm J}}}
\newcommand{\Mp}{\ensuremath{M_p}}
\newcommand{\Ms}{\ensuremath{M_\star}}
\newcommand{\Rs}{\ensuremath{R_\star}}
\newcommand{\Tshift}{\ensuremath{T_{\rm shift}}}

\DeclareGraphicsExtensions{.png,.pdf}

\defcitealias{Weinberg:2012}{WAQB}
\defcitealias{Essick:2016}{EW16}

\begin{document}

\title{\bf \large Orbital Decay of Hot Jupiters due to Weakly Nonlinear Tidal Dissipation}

\correspondingauthor{Nevin N. Weinberg}
\email{nevin@uta.edu}

\author[0000-0001-9194-2084]{Nevin N. Weinberg}
\affiliation{Department of Physics, University of Texas at Arlington, Arlington, TX 76019, USA}

\author{Niyousha Davachi}
\affiliation{Department of Physics, University of Texas at Arlington, Arlington, TX 76019, USA}

\author[0000-0001-8196-9267]{Reed Essick}
\affiliation{Canadian Institute for Theoretical Astrophysics, University of Toronto, M5S 3H8 Ontario, Canada}

\author[0000-0002-6011-6190]{Hang Yu}
\affiliation{Kavli Institute for Theoretical Physics, University of California at Santa Barbara, Santa Barbara, CA 93106, USA}

\author[0000-0001-5611-1349]{Phil Arras}
\affiliation{Department of Astronomy, University of Virginia, P.O. Box 400325, Charlottesville, VA 22904, USA}

\author[0000-0003-1950-448X]{Brent Belland}
\affiliation{California Institute of Technology, 1200 East California Boulevard, Pasadena, CA 91125, USA}

\shorttitle{Orbital decay of hot Jupiters}

\begin{abstract}

We study tidal dissipation in hot Jupiter host stars due to the nonlinear damping of tidally driven $g$-modes, extending the calculations of \citet{Essick:2016} to a wide variety of non-solar type hosts.  This process causes the planet's orbit to decay and has potentially important consequences for the  evolution and fate of hot Jupiters.  Previous studies either only accounted for linear dissipation processes or assumed that the resonantly excited primary mode becomes strongly nonlinear and breaks as it approaches the stellar center.  However, the great majority of hot Jupiter systems are in the weakly nonlinear regime in which the primary mode does not break but instead excites a sea of secondary modes via three-mode interactions.  We simulate these nonlinear interactions and calculate the net mode dissipation for stars that range in mass from $0.5 M_\sun \le \Ms \le 2.0 M_\sun$ and in age from the early main sequence to the subgiant phase.  For stars with $\Ms \lesssim 1.0 M_\sun$ of nearly any age, we find that the orbital decay time is $\lesssim 100 \trm{ Myr}$ for orbital periods $P_{\rm orb} \lesssim 1 \trm{ day}$.  For $\Ms \gtrsim 1.2 M_\sun$,  the orbital decay time only becomes short on the subgiant branch, where it can be $\lesssim 10 \trm{ Myr}$ for  $P_{\rm orb} \lesssim 2 \trm{ days}$ and result in significant transit time shifts.  We discuss these results in the context of known hot Jupiter systems and examine the prospects for detecting their orbital decay with transit timing measurements.
\end{abstract}


\section{\bf I\lowercase{ntroduction}}
\label{sec:intro}

The orbits of hot Jupiters decay over time due to the tide-induced transfer of energy and angular momentum from the orbit to the host star. The orbital decay rate depends on the efficiency of tidal dissipation within the star  and is sensitive to its structure and evolutionary state.  The rate can therefore be a strong function of not just orbital period, but also stellar mass and age (for a  review of tidal dissipation processes in stars and giant planets, see \citealt{Ogilvie:2014}).

The most direct observational evidence of hot Jupiter orbital decay comes from the measured transit time shifts of  WASP-12b  \citep{Maciejewski:2016, Patra:2017,Yee:2020}  and Kepler-1658b \citep{Vissapragada:2022}.   A number of studies also find evidence from the statistical analysis of hot Jupiter populations   \citep{Jackson:2009,  Teitler:2014, Penev:2018, Hamer:2019}. For example, \citet{Jackson:2009} find that older planets tend to be farther from their hosts than younger planets, which they argue is evidence for the ongoing destruction of planets by tides.  Using the measured Galactic velocity dispersion, \citet{Hamer:2019}  show that  hot Jupiter host stars are preferentially younger than a matched sample of field stars, which can be explained if the planets are destroyed while the hosts are on the main sequence. \citet{McQuillan:2013} find a dearth of close-in planets orbiting rapidly rotating stars, which \citet{Teitler:2014}  attribute to tidal ingestion of giant planets. The recently reported infrared transient  ZTF SLRN-2020 appears to capture the moments of a planet's ingestion by a main sequence or early  subgiant branch star with mass around $0.8-1.5 M_\sun$, and could be the culmination of tide-induced orbital decay \citep{De:2023}.

The dominant source of dissipation in most hot Jupiter host stars is damping of the resonantly excited internal gravity waves that comprise the dynamical tide \citep{Barker:2010, Ivanov:2013, Essick:2016, Barker:2020}. In stars with thick outer convective envelopes, an internal gravity wave is excited near the radiative-convective boundary and propagates inwards towards the core. As the wave approaches the stellar center, its  amplitudes grows due to geometric focusing and it can become nonlinear. If the amplitude of the wave is not too large, it reflects at an inner turning point and forms a standing wave, i.e., a $g$-mode.

Most studies either ignore nonlinearities and treat the wave as linear throughout the star or they take the other extreme and assume it becomes strongly nonlinear and undergoes wave breaking in the core. However, as we show (see also \citealt{Barker:2020}), most hot Jupiter systems are in an intermediate regime where the wave is weakly nonlinear.  In this regime, the wave excites a sea of secondary waves through nonlinear wave-wave interactions. Since the secondary waves have much shorter wavelengths than the primary wave, they have much larger damping rates. Treating the primary wave as linear is therefore not only invalid, it can greatly underestimate the efficiency of tidal dissipation.  On the other hand, treating it as  strongly nonlinear overestimates the efficiency because it assumes the wave transfers all of its energy and angular momentum on its first journey into the stellar center.  In the weakly nonlinear regime, by contrast, the primary wave deposits only a fraction of its energy and angular momentum each journey. The value of the deposited fraction depends on the detailed interaction between the primary and secondary waves and determines the rate of tidal dissipation. 

The first calculation of hot Jupiter orbital decay in the weakly nonlinear regime was by \citeauthor{Essick:2016} (2016; hereafter \citetalias{Essick:2016}).  They assumed a solar-type host star and considered a range of planet masses $\Mp$ and orbital periods $\Porb$.  For a solar-type host, the tide is in the weakly nonlinear regime for  $\Mp \lesssim 3.6 \Mj \left(\Porb/\trm{day}\right)^{-0.1}$;  above this planet mass the waves are strongly nonlinear \citep{Barker:2011a}.  By solving the dynamics of large networks of nonlinearly interacting waves, \citetalias{Essick:2016} calculated the weakly nonlinear tidal dissipation and found  a stellar tidal quality factor $\Q \simeq 3\times10^5 \left(\Mp/\Mj\right)^{0.5}(\Porb/\trm{day})^{2.4}$.  The dissipation can thus be highly efficient, causing hot Jupiters with solar-type hosts and $\Porb \lesssim 2 \trm{ days}$ to decay on timescales that are small compared to the main-sequence lifetime of their host star. They found that the decrease in $\Porb$ associated with these  short decay times could produce detectable transit timing variations.

\begin{figure}[t]
\centering
\includegraphics[width=1.0\linewidth]{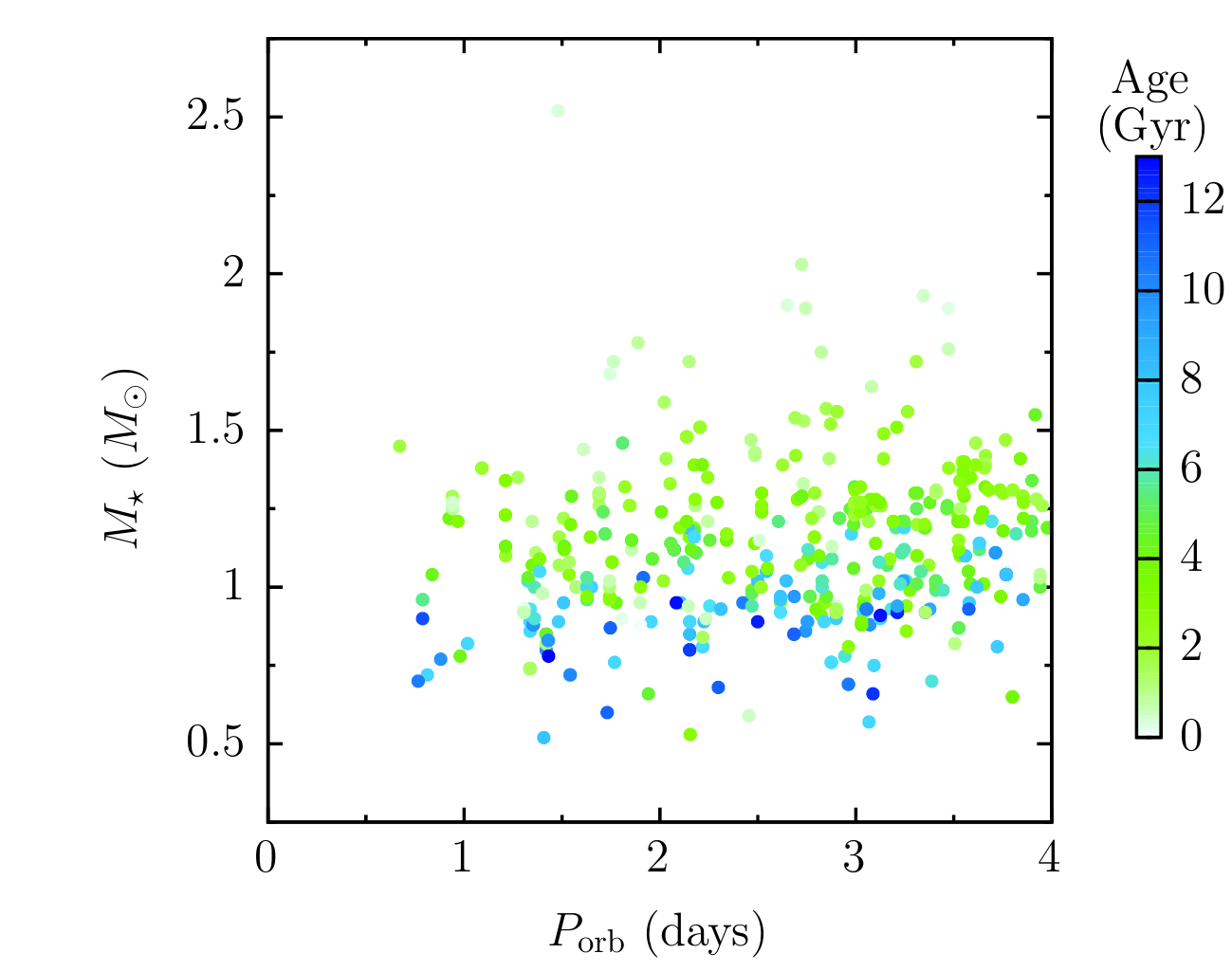}
\caption{Distribution of hot Jupiters in the plane of stellar mass $\Ms$ and orbital period $\Porb$ for planets with mass $M_p \sin i > 0.5 M_J$, where $i$ is the orbital inclination.  The color scale shows the best-fit stellar ages  in Gyrs. 
Data is from the NASA Exoplanet Archive and only stars with reported ages are shown.
\label{fig:hJ_histogram}}
\end{figure}

\citetalias{Essick:2016} only considered hot Jupiters orbiting a solar-type star.  However, as we show in Figure~\ref{fig:hJ_histogram}, there are many hot Jupiters orbiting non-solar type stars. The results of \citetalias{Essick:2016} do not inform these other systems since the efficiency of tidal dissipation  is often sensitive to stellar structure and therefore stellar mass $\Ms$ and age.  For example, on the main sequence the mass of the convective envelope rapidly decreases with increasing $\Ms$, and a convective core appears above $\Ms \approx 1.2 M_\sun$. While the former effect may lead to only modest changes in dissipation rate for $\Ms \lesssim 1.2 M_\sun$, the appearance of a convective core may drastically decrease the degree of wave nonlinearity, causing a sudden decrease in tidal dissipation relative to slightly lower mass stars.  As we show, the efficiency of tidal dissipation can also be sensitive to stellar age, especially as stars begin to evolve up the subgiant branch.  

The paper is organized as follows.  In Section~\ref{sec:formalism}, we describe the formalism we use to study weakly nonlinear tidal dissipation, including the equation of motion for our mode decomposition.  In Section~\ref{sec:properties}, we present our stellar models and their linear and nonlinear mode properties. In Section~\ref{sec:networks}, we describe our method for building and integrating  networks of nonlinearly interacting modes.  In Section~\ref{sec:results}, we present the main results of our calculations, including how the orbital decay rate and transit time shift depend on $\Ms$, stellar age, $\Mp$, and $\Porb$. In Section~\ref{sec:implications}, we discuss the implications of our orbital decay calculations for known hot Jupiter systems. We summarize and conclude in Section~\ref{sec:conclusions}.

\section{\bf F\lowercase{ormalism for} W\lowercase{eakly} N\lowercase{onlinear} T\lowercase{ides}}
\label{sec:formalism}

In hot Jupiter systems, the tide raised by the planet excites high-order $g$-modes in the host star through linear and, we assume, weakly nonlinear forces.   Since the planet's orbital period will in general be much shorter than the rotational period of the star, the $g$-modes are not strongly modified by Coriolis forces and we therefore neglect the star's rotation.\footnote{Hot stars on the main sequence with $\Ms \gtrsim 1.2 M_\sun$ and effective temperatures $T_{\rm eff} \ge 6250\trm{ K}$ (i.e., above the Kraft break; \citealt{Kraft:1967}) can rotate with spin periods of a few days, comparable to $\Porb$ of hot Jupiters.  However, we do not include such stars in our study since their tidal dissipation rate is  negligible owing to the presence of a convective core, as explained in Section~\ref{sec:models}.  The stars with $\Ms\ge 1.2 M_\sun$ we consider are all on the subgiant branch and are observed to have spin periods $\gtrsim 10\trm{ days}$ \citep{Avallone:2022}.}  We also assume that the planet's orbit is circular, as is the case for most of the observed hot Jupiters.\footnote{https://exoplanetarchive.ipac.caltech.edu/} The planet's rotation should then be synchronous with the orbit \citep{Storch:2014,Barker:2014} and thus there is no tidal dissipation within the planet.  We study this problem using the formalism developed in \citeauthor{Weinberg:12} (2012; hereafter \citetalias{Weinberg:2012}; see also \citealt{Schenk:2001,VanHoolst:1994}). Here we summarize the method and refer the reader to  \citetalias{Weinberg:2012} for a more detailed discussion.

\subsection{Mode amplitude equation}

The position $\gv{x}'$ of a fluid element in the perturbed star at time $t$ is related to its position $\gv{x}$ in the unperturbed star by $\gv{x}'=\gv{x} +  \gv{\xi}(\gv{x},t)$,  where $\gv{\xi}(\gv{x},t)$ is the Lagrangian displacement vector. 
The equation of motion for $\gv{\xi}$ to lowest nonlinear order is 
\begin{equation}\label{eq:eq_of_motion}
	\rho \ddot{\gv{\xi}} = \gv{f}_1\left[\gv{\xi}\right]+\gv{f}_2\left[\gv{\xi},\gv{\xi}\right]+\rho \gv{a}_{\rm tide},
\end{equation}
\noindent
where $\rho$ is the background density, $\gv{f}_1$ and $\gv{f}_2$ are the linear and leading-order nonlinear restoring forces, 
\begin{equation}
	\gv{a}_{\rm tide}=-\grad U - \left(\gv{\xi}\cdot\grad\right)\grad U
\end{equation}
\noindent
is the tidal acceleration, and $U$ is the tidal potential. We include only the dominant $l=2$ tidal harmonic and since we assume that the orbit is circular, 
\begin{equation}\label{eq:tidal_potential}
	U(\gv{x},t)=- \epsilon \omega_\star^2 r^2 \sum_{m=-2}^2 W_{2m} Y_{2m}(\theta, \phi)e^{-im\Omega t}
\end{equation}
\noindent
in a spherical coordinate system $(r, \theta, \phi)$ centered on the star.  Here $\epsilon = (\Mp/\Ms)(\Rs/a)^3$ is the tidal strength, $\omega_\star=(G\Ms/\Rs^3)^{1/2}$ 
is the dynamical frequency of a star with mass $\Ms$ and radius $\Rs$,  $a$ and $\Omega = \sqrt{G(\Ms+\Mp)/a^3}$ are the orbital semi-major axis and frequency, and $W_{20}=-(\pi/5)^{1/2}$, $W_{2\pm2}=(3\pi/10)^{1/2}$, $W_{2\pm1}=0$.
We use the linear eigenmodes to expand the six-dimensional phase-space vector as \citep{Schenk:02}
\begin{equation}\label{eq:eig_expansion}
	\begin{bmatrix} \gv{\xi}(\gv{x},t) \\ \dot{\gv{\xi}}(\gv{x},t) \end{bmatrix}
	= 
	\sum_{a} q_a(t) \begin{bmatrix} \gv{\xi}_a(\gv{x}) \\ -i\omega_a \gv{\xi}_a(\gv{x}) \end{bmatrix},
\end{equation}
\noindent
where each eigenmode is specified by its amplitude $q_a$,
frequency $\omega_a$, and eigenfunction $\gv{\xi}_a$. The sum over $a$ runs over all mode quantum numbers (radial order $n_a$, angular
degree $l_a$, and azimuthal order $m_a$) and frequency signs to allow
both a mode and its complex conjugate. We normalize the eigenmodes as
\begin{equation}
	E_\star \equiv \frac{G\Ms^2}{\Rs}=2\omega_a^2\int d^3x \rho \, \gv{\xi}_{a}^\ast \cdot \gv{\xi}_{a},
\end{equation}
\noindent
so that the energy of a mode is $E_a(t)=q_a^\ast(t) q_a(t) E_\star$.\footnote{The total energy in the modes includes a higher order term  $(1/3) \sum k_{abc}(q_a q_b q_c + \trm{c.c.})$, as does their total dissipation rate $\sum k_{abc} \gamma_a (q_a q_b q_c + \trm{c.c.})$, where the sums are over all indices and c.c. stands for complex conjugate. However, since $\kappa_{abc}$ is much smaller than the inverse of the typical mode amplitudes, these terms are small compared to the lowest order terms and are neglected.}  
Substituting Equation (\ref{eq:eig_expansion}) into Equation (\ref{eq:eq_of_motion}), adding a linear damping term, and using the orthogonality of the eigenmodes leads to a coupled, nonlinear amplitude equation for each mode
\beq
	\dot{q}_a + (i\omega_a + \gamma_a ) q_a =  i\omega_a \left[U_a 
	+\sum_b U_{ab}^{\ast} q_{b}^\ast  + \sum_{bc} \kappa_{abc} q_{b}^\ast q_c^\ast\right],
\label{eq:amp_eqn_orig}
\eeq
where 
\begin{subequations}
	\begin{align}
		U_a(t)                = &-\frac{1}{E_\star}\int d^3x \rho \, \gv{\xi}_a^\ast \cdot \grad U,\label{eq:Ualpha}\\
		U_{ab}(t)         = &-\frac{1}{E_\star}\int d^3x \rho \, \gv{\xi}_a \cdot\left(\gv{\xi}_{b} \cdot \grad\right)\grad U,\\
		\kappa_{abc} = &\frac{1}{E_\star}\int d^3x\, \gv{\xi}_{a}\cdot\gv{f}_2\left[\gv{\xi}_b, \gv{\xi}_c\right].
	\end{align}
\end{subequations}
\noindent
The coefficient $\gamma_a$ is the linear damping rate of the mode, $U_a$ and $U_{a b}$ are the linear and nonlinear tidal coefficients, and $\kappa_{a b c}$ is the three-mode coupling coefficient.   

The nonlinear tide cancels significantly with three-mode coupling to the equilibrium tide such that $U_{ab} + 2\sum_c\kappa_{abc}U_c \simeq 0$ \citepalias{Weinberg:2012}. By treating the cancellation as perfect, we can write Equation~(\ref{eq:amp_eqn_orig}) as
\begin{multline}
	\dot{q}_a + (i\omega_a + \gamma_a ) q_a =  i\omega_a U_a \\
	+i\omega_a \sum_{bc} \kappa_{abc} \left[q_{b}^\ast q_c^\ast - q_b^\ast U_c^\ast - q_c^\ast U_b^\ast\right].
\label{eq:amp_eqn}
\end{multline}
We solve this coupled set of equations to find the amplitude evolution $q_a(t)$ of each mode in our network and from this determine the total tidal dissipation rate  \citepalias{Essick:2016}
\beq
\dot{E}(t) = -2\sum_a \gamma_a E_a(t).
\eeq
In Section~\ref{sec:properties}, we  present the values of $\gamma_a$, $U_a$, and $\kappa_{abc}$ for our stellar models.

\subsection{Parametric instability}

As discussed in \citetalias{Weinberg:2012} and \citetalias{Essick:2016}, in the absence of nonlinear mode coupling the linear tide drives a mode to an energy
\begin{equation}
	E_{\rm lin} = \frac{ \omega_a^2 U_a^2 } {\Delta_a^2 + \gamma_a^2} E_\star,
\label{eq:Elin}
\end{equation}
where $\Delta_a= \omega_a - m_a \Omega$ is the linear detuning.  This `parent' mode is unstable to the parametric instability if there exists a pair of `daughter' modes $b$ and $c$ such that  $E_{ \rm lin} \ga E_{\rm thr}$, where the threshold energy is 
\begin{equation}
	E_{\rm thr} = \frac{1}{4\kappa_{abc}^2}\left(\frac{\gamma_b \gamma_c}{\omega_b \omega_c}\right)\left[1+\left(\frac{\Delta_{abc}}{\gamma_b+\gamma_c}\right)^2\right] E_\star
\label{eq:Ethr}
\end{equation}
\noindent
and $\Delta_{ab c}=\omega_b+\omega_c+m_a\Omega$ is the nonlinear detuning.  
If the parent energy $E_a \gg E_{\rm thr}$, the daughters grow exponentially at a rate
\begin{equation}\label{e:s3}
	\Gamma_{bc} \approx 2 \Omega \left|\kappa_{abc}\right| \sqrt{\frac{E_a}{E_\star}}
\end{equation}
\noindent
while their energy $E_b, E_c \ll E_a$. Eventually, the  system reaches a nonlinear equilibrium as described in Appendix~B of \citetalias{Essick:2016}.  

In the host star of a hot Jupiter, the linearly resonant parents have $E_{\rm lin} \gg E_{\rm thr}$ and they can therefore excite  many daughter pairs \citepalias{Weinberg:2012}.  For example, in a solar model even a $0.1\, \Mj$ companion in a 3 day orbit has  $\sim 10^3$ daughter pairs for which $E_{ \rm lin}>E_{\rm thr}$ (see Figure~1 in \citetalias{Essick:2016}). Many of these daughters are nonlinearly driven to such large amplitudes that they in turn excite many granddaughters to large amplitudes, and the granddaughters excite great-granddaughters, and so on.   The total number of unstable modes is therefore very large and the system's nonlinear equilibrium is much more complicated than the simple, analytic three-mode equilibrium given in  Appendix~B of \citetalias{Essick:2016}.    In Section~\ref{sec:networks}, we describe our procedure for building and integrating these large networks of nonlinearly excited modes. 

\subsection{Orbital decay rate}

Dissipation of the tidally excited modes removes energy from the orbit and as a result the planet inspirals.  As in \citetalias{Essick:2016}, we assume that linear damping of the waves excited within the star is the only dissipation in the system. Although the rotational energies of the star and the synchronized planet increase as the orbit decays, these changes are small compared to the corresponding change in orbital energy. 
Similarly, the energy in the excited stellar modes themselves and in their interaction energy may change with orbital period, but these too are small effects.

\begin{table*}[!t]
\centering
\caption{Properties of the stellar models and their modes. } \label{tab:models}
\begin{threeparttable}[t]
\centering
\begin{tabular}{*{8}{c}}\hline
\colhead{$\Ms$} &
\colhead{$\Rs$} &
\colhead{Age} &
\colhead{$\Delta P$} &
\colhead{$\log_{10} \gamma_0$} &
\colhead{$\log_{10} I_0$} &
\colhead{$\log_{10} \kappa_0$} &
\colhead{$P_{\rm crit}$\tnote{a}} \\
$(M_\sun)$	 	 &  $(R_\sun)$  & (Gyr)   & ($10^3 \trm{ s}$) &  &   &   & (days)  \\ \hline
0.5	 	 &  0.45  &  1.07   &   9.3   &   $-13.2$    &	   $-1.7$	    &    2.2   & 37.5  \\
0.5	 	 &  0.45  &  4.97   &   6.6   &   $-13.1$    &	   $-1.8$	    &    3.3   & 33.6  \\
0.5  	 &  0.46  &  8.97   &   5.7   &   $-13.0$    &	   $-1.9$	    &    3.6   & 31.3  \\ \hline
0.8	 	 &  0.73  &  1.07   &   4.6   &   $-11.8$    &	   $-2.2$	    &    3.0   & 21.2  \\
0.8	 	 &  0.75  &  4.87   &   3.3   &   $-11.7$    &	   $-2.2$	    &    3.7   & 18.2  \\
0.8	 	 &  0.77  &  8.97   &   2.5   &   $-11.4$    &	   $-2.3$	    &    4.0   & 15.1  \\ \hline
1.0	 	 &  0.91  &  1.03   &   3.8   &   $-11.2$    &	   $-2.4$	    &    3.2   & 16.4   \\
1.0	 	 &  1.00  &  4.60   &   2.1   &   $-10.7$    &	   $-2.6$	    &    4.1   & 12.0   \\
1.0	 	 &  1.20  &  8.72   &   1.1   &   $-9.2$	    &	   $-2.8$	    &    4.5   &  5.3   \\ 
1.0	 	 &  1.55  &  11.0   &   0.48  &   $-8.2$	    &    $-3.0$	    &    5.1   &  3.3   \\ \hline
1.2	 	 &  1.50  &  4.00   &   1.2   &   $-8.7$	    &    $-3.3$	    &    4.4   &  4.8  \\ 
1.2	 	 &  4.00  &  6.00   &   0.13  &   $-4.4$	    &    $-3.6$	    &    5.2   &  0.7  \\ \hline
1.5	 	 &  2.21  &  2.03   &   0.89  &   $-7.7$	    &    $-4.9$	    &    4.7   &  3.5  \\
1.5	 	 &  3.34  &  2.50   &   0.15  &   $-5.3$	    &    $-3.3$	    &    4.8   &  0.9  \\ \hline
2.0	 	 &  2.33  &  0.70   &   1.1   &   $-7.1$	    &    $-5.2$	    &    4.6   &  2.6  \\
2.0	 	 &  5.21  &  0.92   &   0.14  &   $-4.2$	    &    $-3.6$	    &    5.4   &  0.7  \\ \hline
\end{tabular}
\begin{tablenotes}
  \centering
  \item[a] The values listed here assume $l_{a} =2$ (see Equation~(\ref{eq:Pcrit})).
\end{tablenotes}
 \end{threeparttable}
\end{table*}

Since $|\dot{E}/E_{\rm orb}| \ll \Omega$, where $E_{\rm orb}=-G\Ms\Mp/2a$ is the orbital energy, we model the back-reaction on the orbit as a steady decrease in $E_{\rm orb}$ of quasi-Keplerian circular orbits. At each orbital period $\Porb=2\pi/\Omega$, the timescale of orbital energy decay is then given by
\begin{equation}\label{eq:instant tau}
	\T(t) = \frac{E_{\rm orb}}{\dot{E} (t)}.
\end{equation}
\noindent
We can compute a corresponding time-averaged decay time
\begin{equation}\label{eq:tau}
	\aveT = \frac{a}{\left|\dot{a}\right|}=\frac{E_{\rm orb}}{\langle \dot{E} \rangle},
\end{equation}
\noindent
where $\langle \dot{E} \rangle$ is the time-averaged energy dissipation rate. In our calculations, we average over timescales of $\sim 10^6 \Porb$, which is much longer than the time it takes for the system to reach a nonlinear equilibrium but much less than the orbital decay timescales. It is common to parameterize \aveT\ in terms of the star's tidal quality factor (\citealt{Goldreich:1966}; see also \citealt{Jackson:2008})
\bea
\Q &=& 
\frac{9}{2}\Omega \left(\frac{\Mp}{\Ms}\right)\left(\frac{\Rs}{a}\right)^5 \aveT
\non &\simeq &
7.5\times10^{6} \left(\frac{\aveT}{\mathrm{Gyr}}\right) \left(\frac{\Mp}{\Mj}\right) 
        \non && \times
\left(\frac{\Ms}{M_\sun}\right)^{-8/3}  \left(\frac{\Rs}{R_\sun}\right)^{5}\left(\frac{\Porb}{\mathrm{day}}\right)^{-13/3} , 
\label{eq:Q}
\eea
where the expression assumes a circular orbit.  

\subsection{Transit time shift}

The decrease in $\Porb$ due to tide-induced orbital decay will cause a planet's observed transits to arrive early.  Over a duration $T_{\rm dur}$, the transit time will shift by an amount \citep{Birkby:2014}
\beq
\Tshift =
\frac{3}{4}\frac{T_{\rm dur}^2}{\aveT}
\simeq
2.4\trm{ s} \left(\frac{T_{\rm dur}}{10\trm{ yr}}\right)^2
 \left(\frac{\aveT}{\trm{Gyr}}\right)^{-1}.
\label{eq:Tshift}
\eeq
Based on the typical uncertainties in transit timing measurements of hot Jupiter systems (see, e.g., \citealt{Patra:2020,Maciejewski:2022}),  detecting a tide-induced shift requires $\Tshift \gtrsim 10-100\trm{ s}$.


\section{\bf S\lowercase{tellar} M\lowercase{odels and their} M\lowercase{ode} P\lowercase{roperties}}
\label{sec:properties}

Section~\ref{sec:models} presents the set of stellar models we use in our calculations.  Section~\ref{sec:wave_breaking} discusses the dependence of mode displacement $\xi_r$ on radius and stellar structure and the conditions under which wave breaking occurs.  Section~\ref{sec:coefficients} describes our calculations of the various  coefficients in the amplitude equation and presents their values as a function of the stellar and mode properties.

\subsection{Stellar models}
\label{sec:models}

As we showed in Figure~\ref{fig:hJ_histogram},  most of the hot Jupiter host stars are in the mass range $0.5 \lesssim M_\star/M_\sun \lesssim 2.0$ and span a range of ages (the age is often quite uncertain).  Motivated by these observations, we use the \texttt{MESA} stellar evolution code \citep{Paxton:11,Paxton:13,Paxton:15,Paxton:18,Paxton:19,Jermyn:2022}  to construct stellar models with $M_\star$ and age as listed in Table~\ref{tab:models}.  The key parameters of the \texttt{MESA} inlist files we use to build the models are provided in Appendix~\ref{sec:mesa_inlist}.

Our models range in age from the early main sequence through the subgiant branch, and all have a radiative interior and a convective envelope. For $M_\star \ge 1.2 M_\sun$, we only consider stars on the subgiant branch; nonlinear mode coupling of $g$-modes  is negligible in the interior of pre-main sequence stars or main sequence stars with $M_\star \gtrsim 1.2 M_\sun$.  This is because they have convective cores and their $g$-modes therefore do not  steepen sufficiently, as explained below, before reflecting at the radiative-convective boundary.

\subsection{Wave steepening and breaking}
\label{sec:wave_breaking}

High-order $g$-modes are restored by buoyancy and propagate between  inner and outer turning points determined by the locations at which $\omega_a\simeq N(r)$, where $N$ is the  Brunt-V\"ais\"al\"a buoyancy frequency \citep{Aerts:2010}.    Their inner turning points, $r_{\rm inner}$, are very close to the stellar center and their outer turning points are near the radiative-convective interface $r_c$.

The mode displacements $\xi_r$ steepen towards the stellar center due to geometric focusing and conservation of wave flux \citep{Goodman:1998, Barker:2010}.   Their maximum displacement peaks at $r_{\rm inner}$ and scales
as $\xi_{r, \rm max} \propto r_{\rm inner}^{-2}$. As a result, nearly all the nonlinear mode coupling takes place near the inner turning points of the modes (see Appendix A in \citetalias{Weinberg:2012}).

If the initial amplitude of the linearly resonant $g$-mode is sufficiently large, it can become strongly nonlinear ($k_r \xi_r \gtrsim 1$, where $k_r$ is the radial wavenumber) before reaching $r_{\rm inner}$.  This will cause it to overturn the stratification and break rather than reflect. It should then be treated as a traveling wave rather than a global standing wave. For a given stellar model, the critical initial amplitude $\xi_r$ for wave breaking depends linearly on planet mass and only weakly on orbital period ($\Porb^{1/6}$).  \citet{Barker:2020} calculates the critical planet mass $M_{p, \rm crit}$ above which the $g$-mode will break for a range of stellar models. By comparing with his  Figure~9, we see that all of our main sequence models have $M_{p, \rm crit} > 3.0 \Mj$ and thus their $g$-modes are comfortably in the weakly nonlinear regime (the closest is the solar model which has $M_{p, \rm crit} \approx 3.3 \Mj$; the rest have much larger $M_{p, \rm crit}$). 

The value of $M_{p, \rm crit}$ decreases dramatically as a star evolves off the main sequence \citep{Barker:2020}. This is because as the star evolves its core contracts and the core's nearly constant gradient of $N\simeq Cr$  subsequently increases.  Since $r_{\rm inner}$ of the resonantly excited $g$-mode  is located where $2\Omega\simeq N$, i.e., $r_{\rm inner}\simeq 2\Omega/C$, as the core contracts and $C$ increases, the mode propagates to smaller $r_{\rm inner}$. As a result,  $k_r \xi_{r,\rm max}$ increases, all else being equal. 

This effect can be important for our models with $\Ms \ge 1.2 M_\sun$, which, as explained in Section~\ref{sec:models}, are all on the subgiant branch.  There are two models for each mass in this range, a less evolved and a slightly more evolved  subgiant star.
In the latter, the linearly resonant $g$-modes are close to the wave breaking limit 
for typical hot Jupiter masses \citep{Sun:2018,Barker:2020}.  These stars thus represent the transition between the weakly and strongly nonlinear regimes.  We do not consider even more evolved subgiant stars or stars on the red giant branch since the efficiency of their tidal dissipation is instead found by analyzing the dynamical tide in the traveling wave regime (see, e.g., \citealt{Barker:2010,Barker:2011b,Weinberg:2017,Sun:2018,Barker:2020}).

\subsection{Values of the coefficients in the amplitude equation}\label{sec:coefficients}

\begin{figure}
\centering
\includegraphics[width=0.85\linewidth]{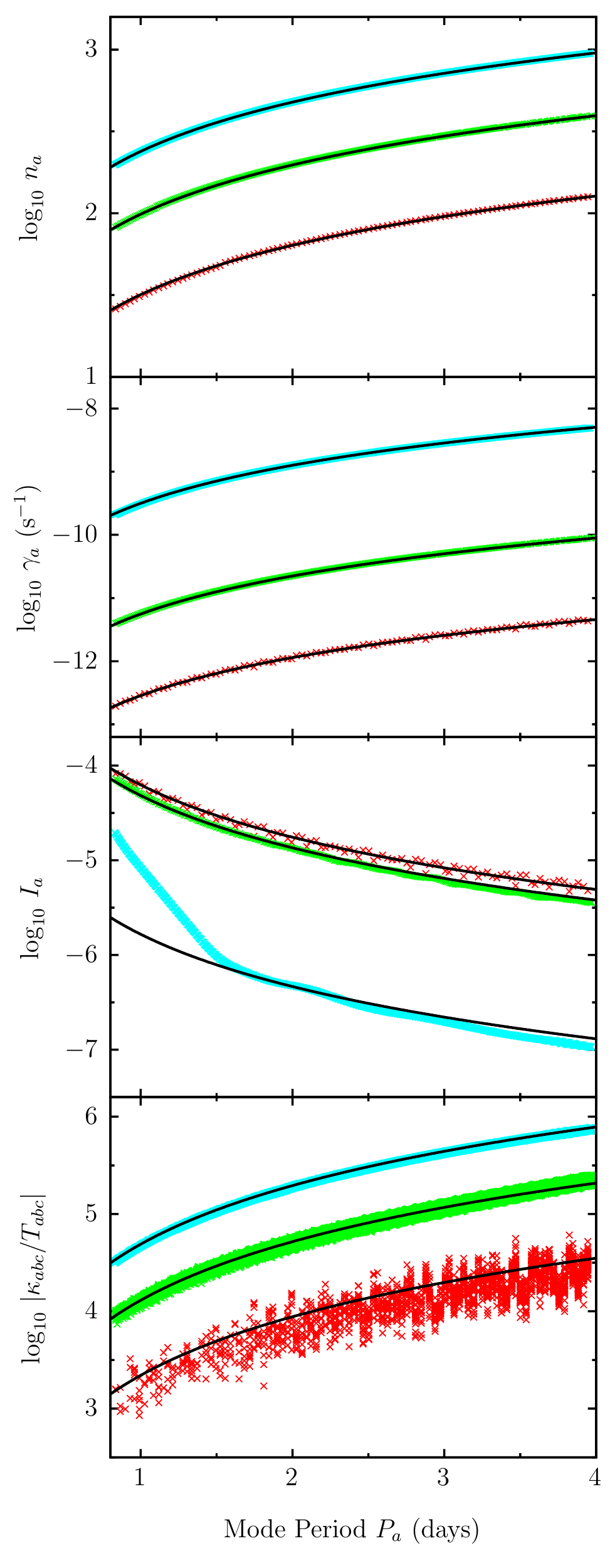}
\caption{From top to bottom panel: Mode radial order $n_a$, damping rate $\gamma_a$, overlap integral $I_a$, and coupling coefficient $\kappa_{abc}/ T_{abc}$ as a function of mode period $P_a$ for $l=2$ modes (see text for details).  Each panel shows three stellar models:  $\left(M/M_{\odot}, \trm{ Age/Gyr}\right)=(0.5, 4.97)$ in red, $(1.0, 4.60)$ in green, and $(1.5, 2.03)$ in cyan.  The solid black lines show the fitted expressions given by Equations~(\ref{eq:omega}), (\ref{eq:gamma}), (\ref{eq:overlap}), and (\ref{eq:kappa}).  
\label{fig:coef_fit_example}}
\end{figure}

Since the $g$-modes we consider are all high-order ($n_a\ga30$), their properties are well-approximated by asymptotic WKB expressions \citep{Aerts:2010}.  For example, their eigenfrequencies are approximately given by
\beq
\omega_a \simeq \frac{2\pi \Lambda_a}{n_a\Delta P},
\label{eq:omega} 
\eeq
where $\Lambda_a = \sqrt{l_a\left(l_a+1\right)}$ and
\beq
\Delta P = \frac{2\pi^2}{\int_0^{r_c} N d\ln r}.
\eeq
We illustrate the accuracy of this scaling in the top panel of Figure~\ref{fig:coef_fit_example} for three of the models: $(\Ms/M_{\sun}, \trm{Age/Gyr})=(0.5,4.97)$, $(1.0, 4.60)$, and $(1.5,2.03)$.  In Table~\ref{tab:models}, we give the value of $\Delta P$ for all our models. 

As we now show, the various coefficients in Equation~(\ref{eq:amp_eqn}) are also all well-fit by power-law expressions.  We use these expressions to build our mode networks since they enable a much faster search for mode triplets with low parametric threshold energies $E_{\rm th}$ (see Section~\ref{sec:networks}).

\subsubsection{Linear damping rate $\gamma_a$}
\label{sec:damping}

The dominant linear damping mechanism of the high-order $g$-modes in all our models is radiative damping, similar to the solar models  shown previously  \citep{Terquem:1998, Goodman:1998}.  We use \texttt{GYRE}'s solution of the non-adiabatic oscillation equations to determine the radiative damping  rate $\gamma_a^{(\mathrm{rad})}$. We also calculate the damping due to turbulent viscosity in  convection zones, which we find  by computing
\begin{align}
    \gamma_a^{(\mathrm{turb})}=\frac{\omega_a^2}{E_\star}\int \mathrm{d}r \, \rho r^2 \nu_{\rm turb} F(r), 
\end{align}
where $F(r)$ is given by the expression derived in \citet{Higgins:68} and  depends on $\gv{\xi}_a(\gv{x})$. The  turbulent effective viscosity $\nu_{\rm turb}$ is a function of the ratio of the convective turnover frequency (provided by \texttt{MESA}) to the mode frequency and decreases as this ratio decreases.  We calculate $\nu_{\rm turb}$ using the power-law expression given in \citet{Duguid:20} from a fit to their numerical simulations. We find that for the dynamically relevant modes, turbulent damping is always much smaller than radiative damping and $\gamma_a \simeq \gamma_a^{(\mathrm{rad})}$.

As in the case of a solar model \citepalias{Weinberg:2012}, the linear damping rate in all our models is well-fit by the expression 
\beq
\gamma_a = \gamma_0 \Lambda_a^2 \left(\frac{\omega_a}{\omega_\star}\right)^{-2}\omega_\star.
\label{eq:gamma} 
\eeq
This dependence stems from the fact that the radiative diffusion is proportional to the second derivative of the temperature fluctuation, and thus the damping rate of short-wavelength perturbations scales as the square of the radial wavenumber $k_a^2 \propto (\Lambda_a/\omega_a)^2$. 

We illustrate the accuracy of this scaling in  Figure~\ref{fig:coef_fit_example} for three of the models (second panel from the top) and in Table~\ref{tab:models} we give the value of $\gamma_0$ for all our models.  We find that $\gamma_0$ increases dramatically with increasing $\Ms$ and stellar age.

Damping attenuates a mode's amplitude by $\exp(-\alpha)$ during a round-trip between the turning points,  where
\beq
\alpha \simeq \gamma_a t_{\rm group} \simeq \frac{4\pi^2 \gamma_0 \Lambda_a^3}{\Delta P \omega_\star}\left(\frac{\omega_a}{\omega_\star}\right)^{-4}
\label{eq:alpha}
\eeq
and $t_{\rm group} = 2 \int_0^{r_c}  \left|v_{\rm group}\right|^{-1} dr$ is the round-trip group travel time of a mode with group velocity $v_{\rm group} =\partial \omega_a / \partial k_a$. The second equality in Equation~(\ref{eq:alpha}) follows from the WKB dispersion relation for $g$-modes (Equation~(\ref{eq:omega})) and the scaling relation for $\gamma_a$ (Equation~(\ref{eq:gamma})).  If $\alpha \gtrsim 1$, a significant fraction of the mode's energy is lost during a round trip and  the mode is effectively a traveling wave (see, e.g.,  \citealt{Goodman:1998,Burkart:13, Sun:2018}).  In that case, our treatment of nonlinear mode coupling, which assumes the modes are global standing waves, is no longer valid.  By Equation~(\ref{eq:alpha}), we see that $\alpha>1$ if the mode period $P_a=2\pi /\omega_a$ exceeds a critical value
\beq
P_{\mathrm{crit}} = \left(\frac{\Delta P}{2\pi \gamma_0}\right)^{1/4}\left(\frac{2\pi}{\Lambda_a \omega_\star}\right)^{3/4}.
\label{eq:Pcrit}
\eeq

In Table~\ref{tab:models} we give the values of $P_{\rm crit}$ for each of our models.  For all the $M_\star \le 1.0 M_\sun$ models and for the early subgiant models with $M_\star \ge 1.2 M_\sun$, the $l_a=2$ values are  in the range $P_{\rm crit} \simeq [3, 40]\trm{ days}$.  Thus, for these models, the linearly  resonant parents ($P_a \simeq P_{\rm orb} /2$) are comfortably in the standing wave regime.  For the older subgiant models with $M_\star \ge 1.2 M_\sun$, this is only true for hot Jupiter systems with $P_{\rm orb} \lesssim 1.5 \trm{ days}$; at longer periods, the parent modes are in the traveling wave regime.  However, we will see  that even if we continue to treat them as standing waves, the mode network calculation yields a tidal dissipation rate that is close to that found with a traveling wave treatment.  

The daughter and granddaughter modes in our networks have periods that are up to a few times larger than the parent modes. Moreover,  they can have $l_a > 2$.  These modes are therefore more likely to be in the traveling wave regime, especially for the older and larger $M_\star$ models.  For such modes, a proper treatment of their nonlinear interactions may require a hybrid formulation in which the parent mode is a standing wave and the secondary modes are traveling waves.  Such a formalism has not been developed, as far as we know, and is left to future work.  We suspect that since the nonlinear mode coupling happens very close to the stellar center (near the inner turning point of the modes) whereas the bulk of the linear dissipation occurs at considerably larger radii, the tidal dissipation rate due to the nonlinear excitation of secondary traveling waves may not be very different from that found here.

\subsubsection{Linear tidal coefficient $U_a$}

By plugging Equation (\ref{eq:tidal_potential}) into Equation (\ref{eq:Ualpha}), we can express the linear tidal coefficient $U_a$ in terms of the dimensionless linear overlap integral
\beq
I_a  = \frac{1}{\Ms \Rs^2}\int d^3x \rho\, \gv{\xi}_{a}^\ast \cdot\grad\left(r^2 Y_{2m}\right).
 \label{eq:forcing}
 \eeq
We evaluate $I_a$ using the  method given by Equation (B23) in \citet{Burkart:13}, which offers good numerical stability even for high-order modes.    As in the solar model (see Figure 11 of \citetalias{Weinberg:2012}), we find that the $l_a=2$ overlap integrals of our models are generally well-fit by
\beq
I_a=I_0\left(\frac{\omega_a}{\omega_\star}\right)^{11/6}.\label{eq:overlap} 
\eeq
We illustrate the accuracy of this scaling in  Figure~\ref{fig:coef_fit_example} for three of the models (third panel from the top) and in Table~\ref{tab:models} we give the value of $I_0$ for all our models.  We find that $I_0$ decreases with increasing $\Ms$ and stellar age (except for the most evolved and massive models, where it increases with age up the subgiant branch).  For our high $\Ms$ subgiant models, $I_a$ can be larger than the scaling at short periods (see, e.g., the $1.5 M_\odot$ model in Figure~\ref{fig:coef_fit_example}). This is because in those models, the mode wavelength becomes larger than the size of the convective envelope at short mode periods and the $\omega_a^{11/6}$ scaling only applies in the opposite limit. For those cases, we use the numerically computed $I_a$ rather than the scaling relation.

In our mode networks, we assume that only linearly resonant modes (parents) have non-zero linear tidal forcing $U_a$. 
This is justified because $U_a$ is much smaller for the daughter modes, granddaughter modes, etc., and their driving is far off resonance.  It therefore has a negligible secular effect compared to the resonant three-mode interactions.  Ignoring such forcing allows us to adopt the convenient change of coordinates described in \citetalias{Essick:2016} (see Section~\ref{sec:integration_method} below) and significantly speeds up the integration of the amplitude equations.

\subsubsection{Three-mode coupling coefficient}

We calculate the three-mode coupling coefficient $\kappa_{abc}$, which is symmetric under the interchange of mode indices, using Equations A55 through A62 in \citetalias{Weinberg:2012}.  Angular momentum conservation leads to the following angular selection rules for the three modes: (i) $l_a+l_b+l_c$ must be even, (ii) $m_a+m_b+m_c= 0$, and (iii) the triangle inequality, $\left| l_a-l_b\right| \leq l_c \leq l_a + l_b$.
We focus on the parametric instability involving three-mode interactions between a high-order parent $g$-mode and a pair of  high-order daughter $g$-modes whose summed frequency nearly equals the parent's frequency (and similarly, a daughter can couple to pairs of granddaughters, etc.).
For such a triplet, the coupling is strongest in the stellar core,  where the displacements of the modes peak (Section~\ref{sec:wave_breaking}). As in the solar model  (Appendix A in \citetalias{Weinberg:2012}), we find that the coupling coefficients of  our models are well-fit by 
\begin{equation}
\kappa_{abc}=\kappa_0 T_{abc} \left(\frac{P_a}{\textrm{day}}\right)^2,
\label{eq:kappa}	
\end{equation}
where $P_a$ is the period of the parent mode and $T_{abc}\approx 0.1 -1$ is an angular integral that depends on each mode's $l$ and $m$ and is easily evaluated in terms of  Wigner 3-j symbols.  The coupling occurs mostly near the parent's inner turning point $r_{a, \rm inner}$ and scales as $P_a^2$ because the parent's displacement there varies as $\xi_{r,a} \sim  r_{a, \rm inner}^{-2} \sim P_a^{2}$.  

We illustrate the accuracy of this scaling in  Figure~\ref{fig:coef_fit_example} for three of the models (bottom panel).  In the figure, the triplets  consist of $l_a=l_b=l_c=2$ modes and nearly resonant daughters with detuning $|\omega_a+\omega_b + \omega_c| < |10^{-3} \omega_a|$ and $|n_b - n_c| < |0.8 n_a|$.  The magnitude of $\kappa_{abc}$ varies by factors of order unity depending on the particular daughter pair and is negligible if $|n_b - n_c| \gtrsim |n_a|$ (see Figure~12 in \citetalias{Weinberg:2012}). In Table~\ref{tab:models}, we give the value of $\kappa_0$ for all our models.  We find that it increases somewhat with increasing $\Ms$ and stellar age.

\section{\bf B\lowercase{uilding and} I\lowercase{ntegrating  the} M\lowercase{ode} N\lowercase{etworks}}
\label{sec:networks}

As in the solar model calculations of \citetalias{Essick:2016}, we find that the parent can drive many parametrically unstable daughters to large amplitudes.  These daughters can then drive parametrically unstable granddaughters to large amplitudes, and so on.  The total number of potentially unstable modes and the number of couplings is larger than the number we can realistically simulate, especially in a survey that explores the parameter space of $\Ms$, stellar age, $\Porb$, and $\Mp$.  Fortunately, as \citetalias{Essick:2016} found in the case of a solar-type host and as we illustrate in Section~\ref{sec:results} for non-solar type hosts,  the total tidal dissipation rate can be reliably computed with a mode network that contains only a subset of the potentially unstable modes. 
This requires  adopting  \citetalias{Essick:2016}'s approach to constructing mode networks, which we now summarize.

\subsection{Selecting daughter modes}

The approach in \citetalias{Essick:2016} relies on the fact that despite the many daughter pairs with $E_{\rm thr}<E_{\rm lin}$ (see Equations~(\ref{eq:Ethr}) and (\ref{eq:Elin})), those with the lowest $E_{\rm thr}$  overwhelmingly dominate the dynamics in large multi-mode systems.  By adding pairs with progressively higher $E_{\rm thr}$ to their networks,  \citetalias{Essick:2016}  found that the system converges to a dissipation rate $\dot{E}$ that does not change significantly as  even more modes are added.   This requires including a sufficient number of generations (at least parents, daughters, and granddaughters) in order to obtain convergent results.  Following \citetalias{Essick:2016}, we thus build our mode networks by systematically searching the mode parameter space and constructing, for each generation, a complete list of pairs ranked by $E_{\rm thr}$.

In order to carry out our search, we use the scaling relations for $\omega_a$, $\gamma_a$, and $\kappa_{abc}$ (Equations~(\ref{eq:omega}), (\ref{eq:gamma}),  and (\ref{eq:kappa})) to solve for $E_{\rm thr}$.  
For a given parent mode $a$, we first find the local minima of $E_{\rm thr}$ in the daughter parameter space $\{(n_b$, $l_b$, $m_b$), ($n_c$, $l_c$, $m_c)\}$. Typically, $E_{\rm thr}$ is minimized near where the sum in quadrature of $\Delta_{abc}$ and $\gamma_b + \gamma_c$ is minimized (modulo the angular selection rules and a relatively weak dependence on the angular integral $T_{abc}$).  There is a tradeoff between finding daughters with smaller $\Delta_{abc}$, which favors higher $n$ and thus higher $l$ for a given $\omega$, and daughters with smaller $\gamma$, which favors smaller $l$ since $\gamma \sim l^2$. The regions of small $E_{\rm thr}$ therefore tend to occur where these two countering effects are balanced.  After finding the local minima, we expand our search around the minima and find pairs with progressively higher $E_{\rm thr}$. 
Because $\gamma\sim l^2$, at high enough $l$ the damping dominates detuning, and $E_{\rm thr}$ increases with increasing $l$.  
We truncate our search upon reaching an $l_{\rm max}$ such that $E_{\rm thr}>E_{\rm lin}$ (i.e., a stable triplet).

\begin{figure}
\centering
\includegraphics[width=1.0\linewidth]{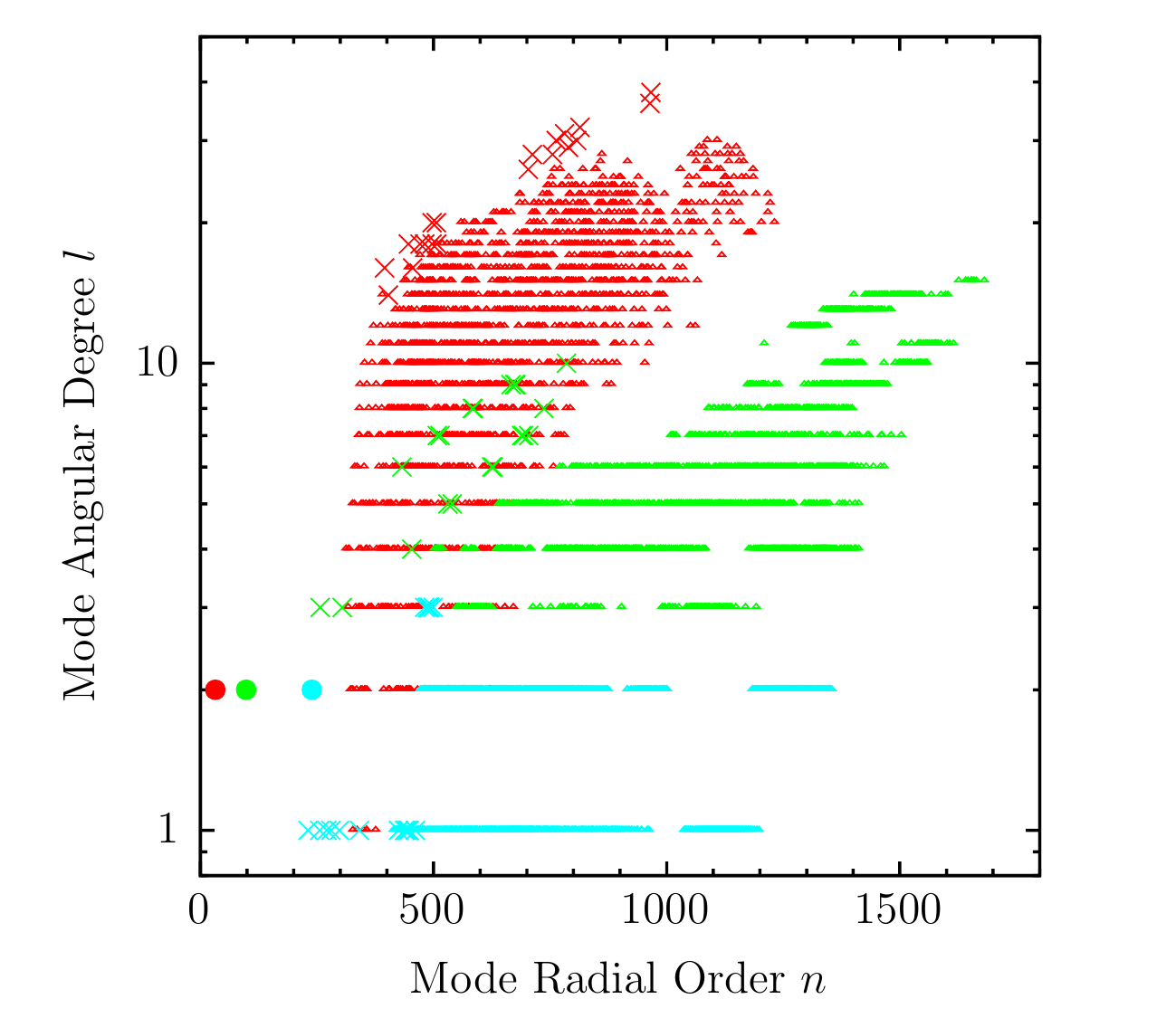}
\caption{Angular degree $l$ and radial order $n$ for each mode in our $P_{\rm orb}=2.0\trm{ day}$ networks for models $\left(\Ms/M_\sun, \trm{ Age/Gyr}\right)=(0.5, 4.97)$ in red, $(1.0, 4.60)$ in green, and $(1.5, 2.03)$ in cyan (the same models and colors as Figure~\ref{fig:coef_fit_example}). The networks consist of one parent mode (solid circles), ten pairs of daughter modes (crosses), and fifty pairs of granddaughter modes per daughter (small triangles).  
\label{fig:network_example}}
\end{figure}

\subsection{Selecting granddaughter modes}

The search for the lowest $E_{\rm thr}$  daughter-granddaughter triplets is carried out in a similar way.  Note that their $E_{\rm thr}$  is generally  smaller than that of  the most unstable parent-daughter triplets.  This is because the mode frequencies decrease with each generation and thus $\kappa_{abc}\propto \omega_a^{-2}$ is larger (Equation~(\ref{eq:kappa})) while $\Delta_{abc}/\omega_a \propto \omega_a^2$ is smaller \citep{Wu:2001}. For a fuller discussion of this point, see Appendix F of \citetalias{Essick:2016}.  Physically, the minimum $E_{\rm thr}$ tends to decrease with each generation for two reasons.  First, lower frequency modes penetrate deeper into the core.  Their peak displacement $\xi_{r, \rm max}$ are therefore greater (for a given mode amplitude) and thus their $\kappa_{abc}$ is larger. Second, such modes are more densely spaced in frequency and therefore can have smaller detunings.   As a result, each generation is more susceptible than its  antecedents to three-mode instabilities.

\subsection{Reference networks}

We find that for parent-daughter (daughter-granddaughter) coupling, the dissipation $\dot{E}$ is overwhelmingly dominated by the $\approx 10$ ($\approx 50$) lowest $E_{\rm thr}$ pairs per parent (daughter), similar to what \citetalias{Essick:2016} found.  Thus, as in \citetalias{Essick:2016}, our reference networks consist of one parent mode, its 10 lowest $E_{\rm thr}$ daughter pairs, and the 50 lowest $E_{\rm thr}$ granddaughter pairs per daughter.   For most networks, this corresponds to $\simeq 20$ daughters and $\simeq 2000$ granddaughters. The exception is the more evolved subgiant models with $\Ms\ge 1.2 M_\sun$ and systems with larger $P_{\rm orb}$.  In those cases, the total number of granddaughters in the network can be several times smaller since many of them couple to more than one daughter  (the number of pairs is still 50).  This is because in those models the modes have much higher damping rates  (see $\gamma_0$ values in Table~\ref{tab:models}), which strongly favors low-$l$ modes and narrows the pool of distinct granddaughters pairs with low $E_{\rm thr}$.

When integrating our networks, we choose the orbital period such that the parent is located midway between linear resonances, i.e., at a resonance trough. Thus, we might set the orbital period to, say, $P_{\rm orb}=1.01\trm{ day}$ rather than $P_{\rm orb}=1.00\trm{ day}$.  We also carry out a restricted set of runs in which the parent is located at a resonance peak.  We will see that for some cases, $\aveT$ is nearly the same regardless of whether the parent is at a resonance trough or peak; for example, this is true of a solar model, as also found by \citetalias{Essick:2016} (see their Figure~2).  In other cases, $\aveT$ is significantly smaller for a parent at a resonance peak.  Such systems will move quickly through resonances and, as a result, the long term decay rate will be close to the off-resonance  $\aveT$ value.

In Figure~\ref{fig:network_example} we show the network structure for three of our stellar models at $P_{\rm orb}=2.0 \trm { day}$;  our other networks have qualitatively similar structure to these.  We see that the daughters and granddaughters with the smallest $E_{\rm th}$ are typically $l\lesssim 10$ modes.  However, for the models with larger $\Delta P$ (i.e., those with smaller $\Ms$ and age), their $l$ values can sometimes be even larger.

\subsection{Integration method}
\label{sec:integration_method}

We integrate each mode's coupled amplitude equation (Equation (\ref{eq:amp_eqn})) using a method similar to \citetalias{Essick:2016}'s (see their Section 3.3 for a discussion).  In particular,  we change coordinates to $x_a= q_a e^{i \omega_a t}$ and  assume that only the linearly resonant parent modes have a non-zero linear tidal forcing $U_a$.  This is a good approximation since the daughters and granddaughters are far from being linearly resonant and they also have much smaller $U_a \propto \omega_a^{11/6}$ (Equation~(\ref{eq:overlap})).  This allows us to greatly speed up the integrations by canceling out the relatively high frequency $\omega_a$ term in the amplitude equation and replacing it with a much lower frequency term proportional to $\Delta_{abc}$. 

We use the \texttt{CVODES} Adams solver from the \texttt{SUNDIALS} package \citep{Hindmarsh:05} and parallelize the computations across multiple CPUs using MPI.   Each  simulation typically takes one to two days to run on a node of a local cluster.

A given simulation run consists of a particular stellar model, orbital period, and planetary mass. For each of our sixteen stellar models (see Table~\ref{tab:models}), we consider $P_{\rm orb}=\{0.5, 1,2,3,4\}\trm{ days}$ and  $M_p=\{0.5,1,2,3\} \, M_{\rm J}$ for a total of $\simeq 300$ runs.

\section{\bf R\lowercase{esults}}
\label{sec:results}

\begin{figure*}
\centering
\includegraphics[width=0.8\linewidth]{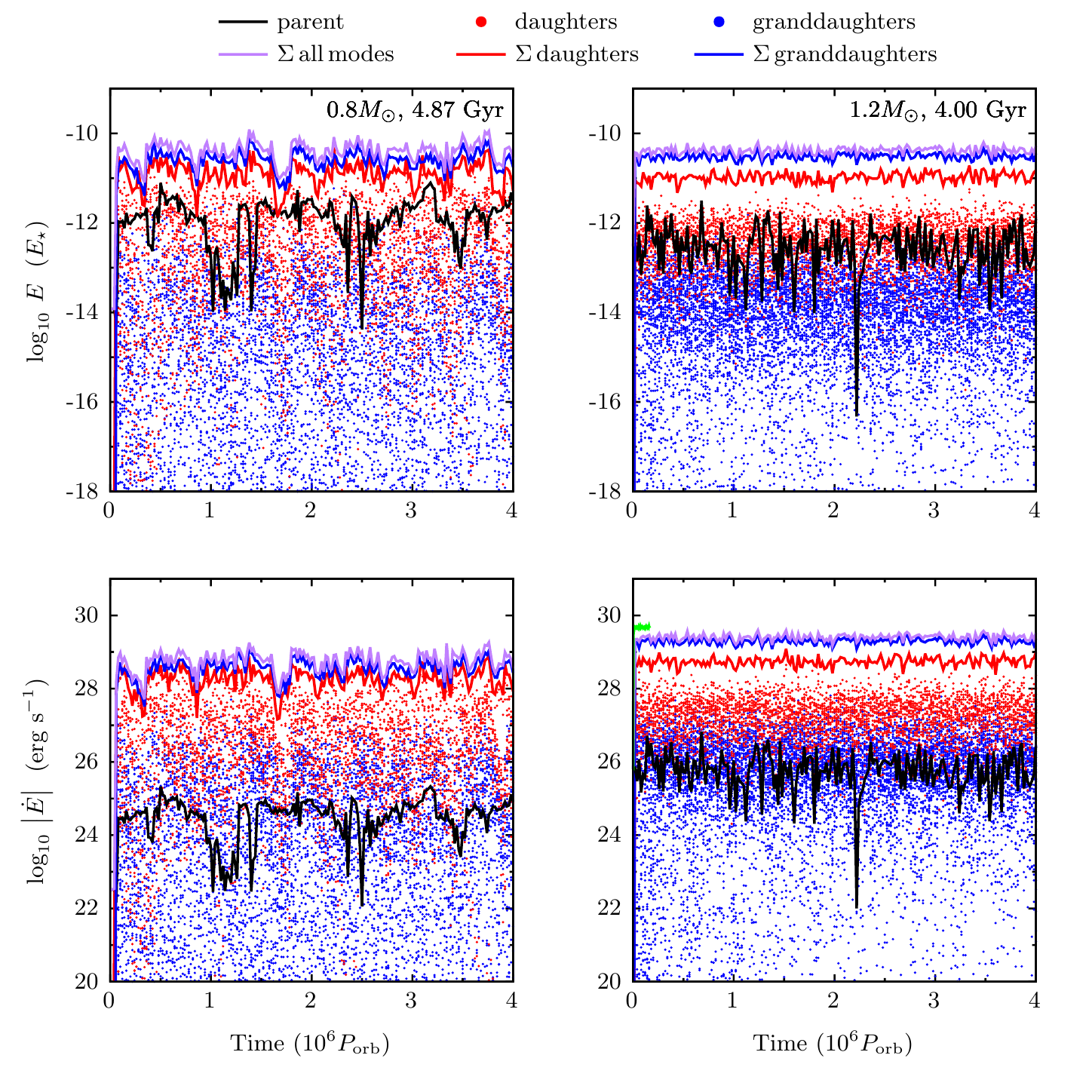}
\caption{Mode energy (top panels) and energy dissipation rate (bottom panels) as a function of time for the planetary parameters $(M_p, P_{\rm orb})=(1.0 M_{\rm J}, 1.0 \trm{ day})$.  The left panels are for the stellar model $(M_\star, \trm{Age})=(0.8 M_\sun, 4.87\trm{ Gyr})$ and the right panels are for $(1.2 M_\sun, 4.00\trm{ Gyr})$. The black lines show the parent mode.  The red points show the twenty daughter modes and the red lines show their cumulative contribution  ($\sum E_a$ and $\sum \dot{E}_a$, where the sums run over only the daughter modes).  The blue points show the $\approx 2000$ granddaughter modes and the blue lines show their cumulative contribution; for clarity, we only plot individual points for one out of every 20 granddaughter modes (many of them also lie below the range of the plotted vertical scale).  The purple lines show the total energy and energy dissipation rate found by summing over all the modes in the network.  The green line in the bottom right panel shows the total energy dissipation rate for a network that includes 100 great-granddaughter pairs per granddaughter.
\label{fig:E_vs_t}}
\end{figure*}

In this section we present the results of integrating the amplitude equation for large sets of nonlinearly coupled modes (Equation~(\ref{eq:amp_eqn})) using the models and mode parameters presented in Section~\ref{sec:properties} and the mode networks built according to the procedure described in Section~\ref{sec:networks}.   In Section~\ref{sec:example_dynamics}, we show examples of the time-dependent mode dynamics for two representative networks. In Sections~\ref{sec:low_mass_results} and \ref{sec:high_mass_results}  we show the full set of results, expressed in terms of the average orbital decay time $\langle \tau\rangle$, the tidal quality factor $Q_\star'$, and the transit time shift $\Tshift$ (Equations~(\ref{eq:tau}), (\ref{eq:Q}), and (\ref{eq:Tshift}), respectively) for the runs from the lower mass stellar models   ($0.5 \le M_\star/M_\sun \le 1.0 $) and the higher mass stellar models  ($1.2 \le M_\star/M_\sun \le 2.0 $), respectively.  Table~\ref{tab:tab_run_data} in  Appendix~\ref{sec:table_of_results} also lists the values of $\aveT$,  $\Q$, and $\Tshift$  from each of the runs.

\begin{figure*}
\centering
\includegraphics[width=0.8\linewidth]{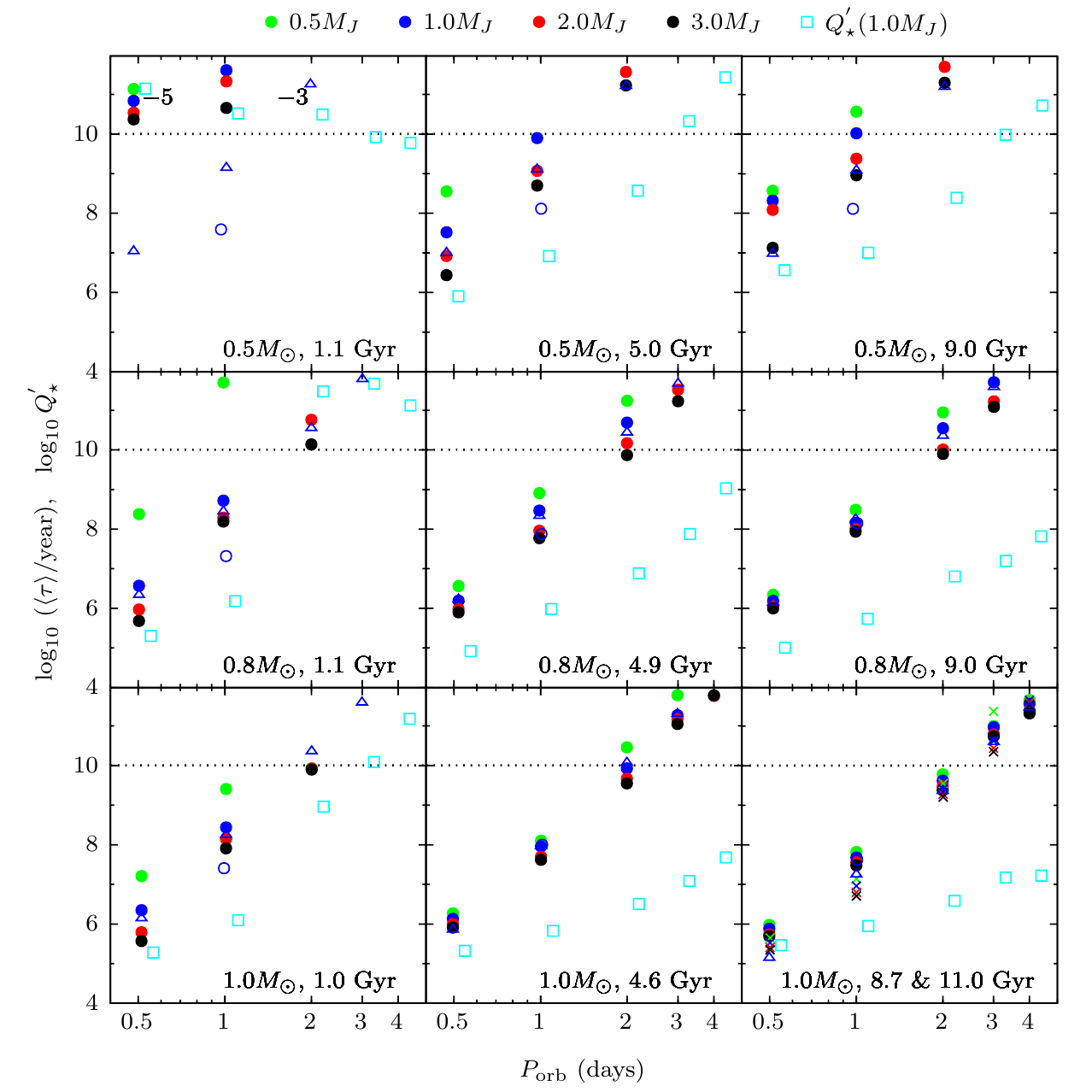}
\caption{Orbital decay time $\langle \tau \rangle$ (solid circles)  as a function of orbital period $P_{\rm orb}$ for the $M_\star=\{0.5, 0.8, 1.0\}M_\sun$ models at different ages, as indicated by the label in each panel.  The green, blue, red, and black solid circles show $\langle \tau \rangle$  for a parent that is off-resonance (i.e., between resonance peaks) and $\Mp=\{0.5,1.0,2.0,3.0\}\Mj$, respectively. The cyan squares show the corresponding stellar tidal quality factor $Q_\star'$ for $\Mp=1.0\Mj$ (the full set of results is given in Table~\ref{tab:tab_run_data}).  The open blue circles show $\langle \tau \rangle$ for a parent on a resonance peak for the case $\Mp=1.0 \, \Mj$, $P_{\rm orb}\simeq 1.0 \trm{ day}$; in four of the panels, the open blue circle blends in with the solid circles since the off-resonance and on-resonance $\langle \tau \rangle$ are similar.  The blue triangles show  $\langle \tau \rangle$ for a parent treated as a traveling wave rather than a standing wave for the case $\Mp=1.0 \, \Mj$.    In the upper left panel, we subtract 5 dex and 3 dex from the off-resonance $\langle \tau \rangle$ and $Q_\star'$, respectively, so that they appear on the plotted scale.  In the bottom right panel, the \texttt{x}'s show  $\langle \tau \rangle$  for the $11.0\trm{ Gyr}$ model using the same color scheme as the solid circles.
\label{fig:tdecay_low_mass}}
\end{figure*}

\begin{figure}
\centering
\includegraphics[width=1.0\linewidth]{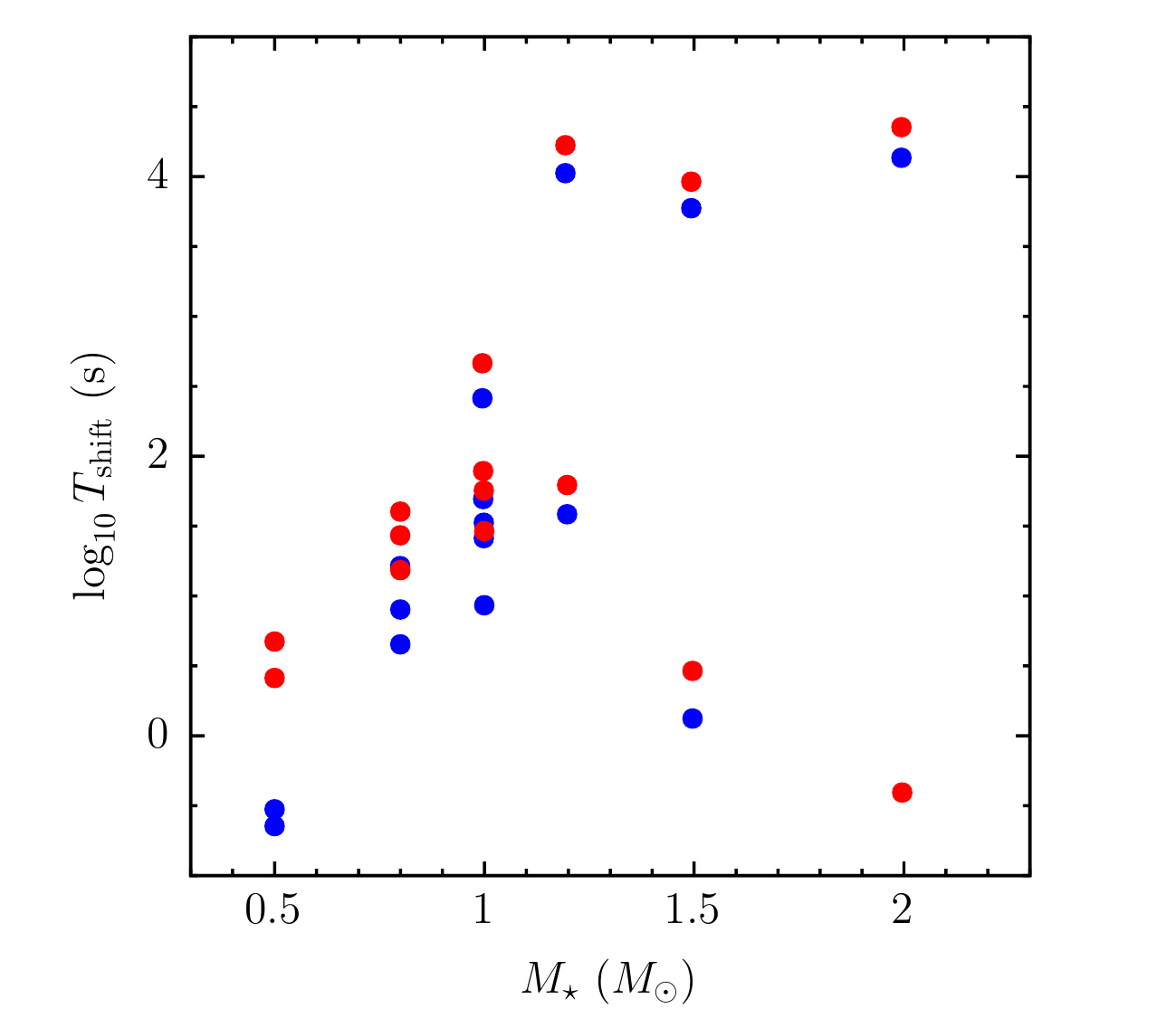}
\caption{Transit time shift $\Tshift$ as a function of stellar mass $\Ms$ for $\Porb=1.0\trm{ day}$ with $\Mp=1.0\Mj$ (blue points) and $\Mp=3.0 \Mj$ (red points) assuming $T_{\rm dur}=10\trm{ yr}$. The ages of the stellar models increase from bottom to top.  In three cases, $T_{\rm shift}$ is below the scale of the plot.  The full set of $T_{\rm shift}$ results are given in Table~\ref{tab:tab_run_data}.
\label{fig:Tshift}}
\end{figure}

\subsection{Example mode dynamics}
\label{sec:example_dynamics}

In Figure~\ref{fig:E_vs_t} we show the time-dependent energy $E_a(t)$ and energy dissipation rate $\dot{E}_a(t)$ of oscillation modes excited in a star due to its tidal interaction with a Jupiter-mass planet on a one day orbit. The left panels show results for a lower mass main sequence star and the right panels show results for a higher mass early subgiant star. In both cases, we see in the top panels that the energy of the linearly resonant parent mode (black line) is similar to the energies of individual daughter modes (red points) and granddaughter modes (blue points).  This is consistent with the notion that upon reaching nonlinear equilibrium, a large system of coupled modes will be in approximate energy equipartition.  

The solid red and blue lines show the summed energy $\sum_a E_a(t)$ of all the daughters and granddaughters, respectively.  We see that the summed energy of all the modes (purple line) is dominated by the contribution of the granddaughters and, to a slightly lesser extent, the daughters and that their contribution is $\sim 10-100$ times larger than the parent's (i.e., while the energy of individual modes tend to be similar across generations, there are many daughters and granddaughters and their summed contribution dominates over the parent's).  Thus, at equilibrium, the energy being pumped into the parent by linear tidal driving is efficiently transferred to the secondary modes.

The bottom panels of Figure~\ref{fig:E_vs_t} show that unlike the energy $E_a(t)$, the energy dissipation rate $\dot{E}_a(t)$ of individual daughter and granddaughter modes is often much larger than that of the parent.  This is because their wavelengths are often much shorter. They thus  have higher linear damping rates $\gamma_a$ and at the same energy are much more dissipative.   The solid lines show that the total energy dissipation rate is dominated by the contribution of the daughters and granddaughters and is $\sim 10^3-10^4$ times larger than the parent's contribution.  We also see that the system reaches a nonlinear equilibrium within $\sim 10^5 P_{\rm orb}$ and, unlike the energy in individual modes, the total dissipation has a fairly small relative standard deviation (1.2 and 0.2 in the left and right panels, respectively).

As described in Section~\ref{sec:networks}, our reference network consists of one parent mode, its 10 lowest $E_{\rm thr}$ daughter pairs, and the 50 lowest $E_{\rm thr}$ granddaughter pairs per daughter; this typically corresponds to $\approx 2000$ total modes.  \citetalias{Essick:2016} found that such a network yields total $\langle\dot{E}\rangle$ comparable to (to within a factor of $\approx 2$) much larger networks that contain more modes and generations.  We carried out a few numerical experiments with larger networks and likewise find that they yield  $\langle\dot{E}\rangle$  values comparable to that of our reference network.  We illustrate this with the green line in the bottom right panel of Figure~\ref{fig:E_vs_t}, which shows the total $\dot{E}$ for the same planetary and stellar parameters as the purple line but for a network that consists of  one parent mode, its 10 lowest $E_{\rm thr}$ daughter pairs, the 50 lowest $E_{\rm thr}$ granddaughter pairs per daughter,  and the 100 lowest $E_{\rm thr}$ great-granddaughter pairs per granddaughter.  The network contains about $2.3\times10^4$ modes and is thus $\simeq 10$ times larger than the reference network  (and comparable in size to the largest networks \citetalias{Essick:2016} ran; see their Figure 3).  Since such a large network runs much slower, we stopped the integration after about $2\times 10^5 P_{\rm orb}$ (the wall time was about ten days).  As the green line shows, this is long enough to see that  $\langle\dot{E}\rangle$ is within a factor of $\simeq 2$ of the reference network.  Similar experiments with a few other networks yield similar results, suggesting that, like in \citetalias{Essick:2016}, the reference network is sufficiently large to yield convergent results.

\subsection{Lower mass stars: $0.5 \le M_\star/M_\sun \le 1.0 $}
\label{sec:low_mass_results}

In Figure~\ref{fig:tdecay_low_mass}, we show the orbital decay time $\aveT$ (solid circles) as a function of $\Porb$ and $\Mp$ for the runs with the $M_\star = \{0.5,0.8,1.0\}M_\sun$ stellar models.  We also show the corresponding stellar tidal quality factor  $Q_\star'$ (open squares) for the $\Mp=1.0 \Mj$ case; in Table~\ref{tab:tab_run_data} of Appendix~\ref{sec:table_of_results}, we list the values  for all the cases.    When computing the time-average, we neglect the early portion of a run (typically the first $10^6$ orbits) to allow transients from the initial conditions to die away and to ensure the system has reached an equilibrium. In Figure~\ref{fig:Tshift}, we show the values of $\Tshift$ for these same models assuming $\Porb=1.0\trm{ day}$  and $\Mp=1.0 \Mj$ or $\Mp=3.0\Mj$ (the full set of values are given in Table~\ref{tab:tab_run_data}).

 As expected, the value of $\aveT$ is most sensitive to $\Porb$; in many cases it has a nearly power-law dependence with $\aveT \propto \Porb^{6.5}$, approximately.  It is also sensitive to the stellar model, with $\aveT$ decreasing as $\Ms$ and stellar age increase.  For all of the $\Ms = 0.8 M_\sun$ and $1.0 M_\sun$ models --- from the early main sequence  to the late main sequence or early subgiant phase --- we find $\aveT \lesssim 10^8\trm{ yr}$ for $\Mp\ge 1.0 \Mj$ and $\Porb \le 1.0 \trm{ day}$.  The corresponding quality factors and transit time shifts are $\Q \approx 10^5 - 10^6$ and $\Tshift \approx 10-100\trm{ s}$.  For the $\Ms = 0.5 M_\sun$ model, $\aveT$ is negligible on the early main sequence (this is largely because $\kappa_0$ is signficantly smaller for this model; see Table~\ref{tab:models}); however, for the older $\Ms=0.5 M_\sun$ models at $\Porb \lesssim 1.0\trm{ day}$ and $\Mp \gtrsim 1.0 \Mj$, we find that $\aveT$ is less than the age of the system, $\Q\lesssim 10^7$, and $\Tshift\gtrsim 10\trm{ s}$.   The value of $\aveT$ tends to be fairly insensitive to planet mass as long as $\Mp \gtrsim 1.0 \Mj$; at smaller planet mass, the nonlinearities start to become less significant and $\aveT$ can become very large, especially for the younger, lower $\Ms$ models.

Our standard calculation assumes the parent is between linear resonance peaks (see Section~\ref{sec:networks}).  To examine how sensitive $\aveT$ is to the parent's degree of  resonance,  for each stellar model we also did a run with a parent located exactly on a resonance peak.  The open blue circles in Figure~\ref{fig:tdecay_low_mass} show the result for the particular case $\Mp=1.0\Mj$ and $\Porb=1.0\trm{ day}$.  For the three $\Ms=0.5 M_\sun$ models and for the young (1.0 Gyr) $\Ms=0.8 M_\sun$ and $1.0 M_\sun$ models, $\aveT$ is significantly smaller for a parent on a resonance peak.  The system will therefore evolve relatively quickly through the resonance peak and ultimately spend most of its time between peaks.    For the other $\Ms=0.8 M_\sun$ and $1.0 M_\sun$ models, $\aveT$ is nearly the same on a resonance peak as it is between peaks (consistent with what \citetalias{Essick:2016} found for the solar model).  Whether on a resonance peak or between, we conclude that the solid circles should be a reasonably accurate estimate of $\aveT$ over long timescales.  A phenomenological model for estimating nonlinear tidal dissipation that can account for differences between on- and off-resonance driving was presented by \citet{Yu:2020} in the context of white dwarf binaries.  In future work, we plan to investigate whether this model can be applied to hot Jupiter systems.

The blue triangles in Figure~\ref{fig:tdecay_low_mass} show $\aveT$ for a parent treated as a traveling wave rather than a standing wave assuming $\Mp=1.0 \Mj$.  The results are found by interpolating across the $Q_{\rm IGW}'$ values given in Figure~8 of \citet{Barker:2020}, applying the $\Porb^{8/3}$ scaling (assuming the tidal forcing frequency is twice the orbital frequency), and using Equation~(\ref{eq:Q}) to convert the results to a decay time $\aveT$.   The parent would be a traveling wave if its amplitude is large enough to undergo wave breaking or if its linear damping is high enough. Although most hot Jupiters are not in this regime (see Sections~\ref{sec:wave_breaking} and \ref{sec:damping}), it is nonetheless interesting to compare the $\aveT$ predictions for the two regimes.   We see that for an off-resonance parent, the traveling wave $\aveT$ is generally smaller then the standing wave $\aveT$, although for the higher $\Ms$ models at short $\Porb$ they can be similar.  By contrast, for an on-resonance parent (see the open blue circles in Figure~\ref{fig:tdecay_low_mass}), the traveling wave $\aveT$ can sometimes be \emph{larger} then the standing wave $\aveT$.\footnote{The traveling wave regime sets a lower limit on $\aveT$ because all the wave flux is lost in a single group travel time.  However, our standing wave calculation  does not ``know'' about this regime.  That said,  many of our models only approach this limit but do not exceed it, which \citetalias{Essick:2016} noticed as well in their solar model calculations (see the discussion in their Appendix F).}  In such cases, this suggests that the parent will be a traveling wave while the system is passing through a resonance and that the correct $\aveT$ during those times is given by the traveling wave value.

\subsection{Higher mass stars: $1.2 \le M_\star /M_\sun \le 2.0 $}
\label{sec:high_mass_results}

\begin{figure}
\centering
\includegraphics[width=1.0\linewidth]{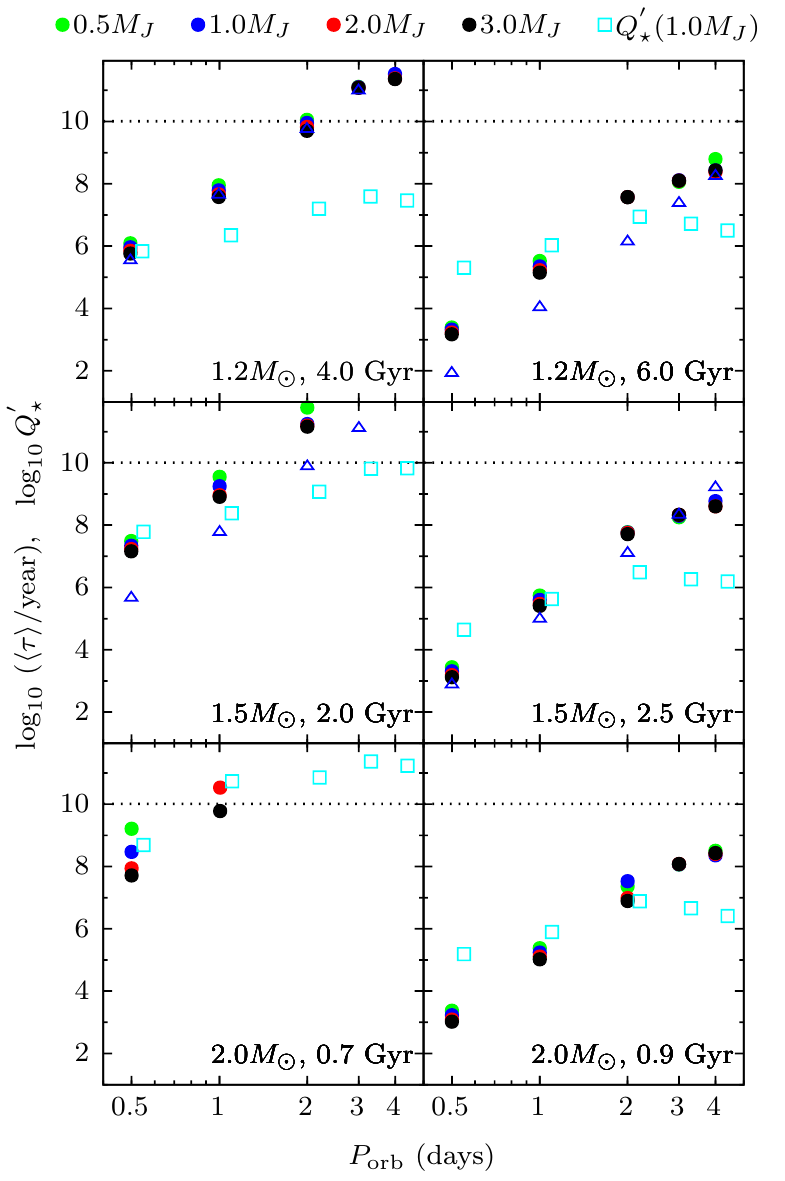}
\caption{Similar to Figure~\ref{fig:tdecay_low_mass} but for the $M_\star=\{1.2, 1.5,2.0\} M_\sun$ models at different ages, as indicated by the label in each panel.  The green, blue, red, and black solid circles show $\langle \tau \rangle$  for an off resonance parent (i.e., between resonance peaks) and $\Mp=\{0.5,1.0,2.0,3.0\}\Mj$, respectively.   The cyan squares show the corresponding stellar tidal quality factor $Q_\star'$ for $\Mp=1.0\Mj$ (the full set of results is given in Table~\ref{tab:tab_run_data}).  The blue triangles in the  $1.2 M_\sun$ and $1.5M_\sun$ panels show  $\langle \tau \rangle$ for a parent treated as a traveling wave rather than a standing wave for the case $\Mp=1.0 \, \Mj$. 
\label{fig:tdecay_high_mass}}
\end{figure}

In Figure~\ref{fig:tdecay_high_mass}  we show $\aveT$ and $\Q$ for the higher mass stellar models: $\Ms=\{1.2, 1.5, 2.0\}M_\sun$.  The corresponding $\Tshift$ values are shown in Figure~\ref{fig:Tshift} and Table~\ref{tab:tab_run_data}.  These models are all in the subgiant phase since, as explained in Section~\ref{sec:models}, main sequence stars with $\Ms \gtrsim 1.2 M_\sun$ have convective cores and nonlinear coupling of $g$-modes is negligible in their interior.  

For each $\Ms$, we find that $\aveT$ decreases significantly as the star evolves from the early to later subgiant phase. We find that at $\Porb \lesssim 1.0\trm{ day}$, the dissipation can be especially efficient in the later subgiant models, with $\aveT \lesssim 10^5\trm{ yr}$, $\Q \lesssim 10^5$ and $\Tshift \gtrsim 10^4 (T_{\rm dur}/10\trm{ yr})^2 \trm{ s}$.  As  Figure~\ref{fig:Tshift} and Table~\ref{tab:tab_run_data} show, the $\Tshift$ of these models are orders of magnitude larger than that of the $\Ms \le 1.0 M_\sun$ main sequence models.  Since $\Tshift\approx 100\trm{ s}$ should be detectable, such systems should produce a detectable transit time shift after only $T_{\rm dur} \approx 1\trm{ yr}$.

The orbital decay is much faster for more evolved subgiants because they have significantly larger $\gamma_0$ and $I_0$, and to a lesser extent, larger $\kappa_0$ (see Table~\ref{tab:models}).  The damping rate is larger because as the stars evolve up the subgiant branch, the core contracts and $C=N/r$ increases.  Since $k_r \propto C/\omega$, at a given $\Porb$ the resonant modes will have shorter wavelengths and thus larger damping rates.  Even at $\Porb\simeq 4\trm{ day}$, the longest orbital period we consider, $\aveT \approx 10^8\trm{ yr}$ for the older subgiant models.  This is comparable to the subgiant evolutionary timescale, and thus at these larger $\Porb$ the changing stellar structure can dictate the planet's decay trajectory (see also \citealt{Sun:2018}).  Interestingly, in those models $\aveT$ is nearly independent of $\Mp$.  Note too that $\aveT$ is also close to the traveling wave value (blue triangles); indeed, for these older stellar models, the critical planet mass for wave breaking is around $1.0 \Mj$ \citep{Sun:2018,Barker:2020} and thus the parent is close to, if not in, the traveling wave regime.

\section{\bf I\lowercase{mplications} \lowercase{for} K\lowercase{nown} H\lowercase{ot} J\lowercase{upiters}}
\label{sec:implications}

We now consider the implications of our orbital decay calculations for known hot Jupiter systems.  In Section~\ref{sec:wasp12}, we discuss the two systems, WASP-12 and Kepler-1658, with observational evidence of tide-induced orbital decay and place them in the context of our results. In Section~\ref{sec:promising}, we discuss the systems that our calculations suggest are especially likely to be undergoing rapid  orbital decay and we recommend be included in campaigns that search for tide-induced transit timing variations.

\subsection{Systems with evidence of orbital decay}
\label{sec:wasp12}

There are two hot Jupiter systems whose transit timing observations indicate that the planet is undergoing orbital decay: WASP-12  \citep{Maciejewski:2016, Patra:2017,Yee:2020}  and Kepler-1658 \citep{Vissapragada:2022}. WASP-12 (Kepler-1658) has an orbital period  $\Porb=1.09\trm{ days}$ ($3.85\trm{ days}$) and a planet mass $\Mp=1.4 \Mj$ ($5.9 \Mj$).  Its orbital period is observed to decrease at a rate $\dot{P}_{\rm orb}=-29\pm 2\trm{ ms yr}^{-1}$ ($-131^{+20}_{-22}\trm{ ms yr}^{-1}$), which corresponds to a decay time $\T=2\Porb/3|\dot{P}_{\rm orb}| \approx 2.2 \trm{ Myr}$ ($1.7\trm{ Myr}$), and  $\Q= 1.8 \times 10^5$ ($ 2.5\times 10^4$). These systems were detected in 2009 \citep{Hebb:2009} and in 2011 \citep{Brown:2011} and thus by Equation~(\ref{eq:Tshift}), $\Tshift \approx 10^3\trm{ s}$ given that it was $T_{\rm dur}\approx 10\trm{ year}$ before evidence of their orbital decay was first reported.

\begin{table}[!t]\centering
\caption{\label{tab:low_mass_data}
Favorable targets for the search for tidal orbital decay for host stars with $\Ms \le 1.1 M_\sun$.  The age column includes the uncertainty, which is symmetric about the best fit if enclosed in parentheses. The fractional uncertainties are generally $\lesssim 0.001\%$ for $\Porb$  and  $\lesssim 10\%$ for $\Ms$ and $\Mp$.\\
{\bf References.}
 (1) \citet{McCormac:2020}; 
 (2) \citet{Bonomo:2017}; 
 (3) \citet{Penev:2016}; 
 (4) \citet{Vines:2019}; 
 (5) \citet{Yee:2023}; 
 (6) \citet{Nielsen:2020}; 
 (7) \citet{Stassun:2017}; 
 (8) \citet{Dai:2017}; 
 (9) \citet{Bouma:2019}; 
 (10) \citet{Oliveira:2019}; 
 (11) \citet{Cortes-Zuleta:2020};  
 (12) \citet{Hellier:2019}; 
  (13) \citet{Spake:2016}; 
(14) \citet{Southworth:2012}. 
}
\begin{tabular}{lccccc}\toprule
&$P_{\rm orb}$ &$\Ms$ &$\Mp$ &Age & \\\cmidrule{2-5}
Hostname &(days) &($M_\sun$) &($\Mj$) &(Gyr) &Refs. \\\midrule
NGTS-10 &0.77 &0.70 &2.16 &$10.4(2.5)$ &(1) \\
WASP-19 &0.79 &0.96 &1.16 &$5.5^{+8.5}_{-4.5}$ &(2) \\
WASP-43 &0.81 &0.72 &2.05 &$7.0(7.0)$ &(2) \\
HATS-18 &0.84 &1.04 &1.98 &$4.2(2.2)$ &(3) \\
NGTS-6 &0.88 &0.77 &1.34 &$9.77^{+0.25}_{-0.54}$ &(4) \\
TOI-1937 A &0.95 &1.07 &2.01 &$3.6^{+3.1}_{-2.3}$ &(5) \\
HIP 65 A &0.98 &0.78 &3.21 &$4.1^{+4.3}_{-2.8}$ &(6) \\
WTS-2 &1.02 &0.82 &1.27 &$7.00^{+6.50}_{-6.40}$ &(2) \\
HAT-P-23 &1.21 &0.58 &1.34 &$4.0(1.0)$ &(7) \\
TrES-3 &1.31 &0.93 &1.80 &$0.90^{+2.80}_{-0.80}$ &(2) \\
HAT-P-36 &1.33 &1.02 &1.85 &$6.6^{+2.9}_{-1.8}$ &(2) \\
Qatar-2 &1.34 &0.74 &2.60 &$1.4(0.3)$ &(8) \\
WASP-4 &1.34 &0.86 &1.19 &$7.0(2.9)$ &(9) \\
HATS-24 &1.35 &1.07 &2.26 &$3.7^{+2.0}_{-1.8}$ &(10) \\
HATS-2 &1.35 &0.88 &1.53 &$9.7(2.9)$ &(2) \\
WASP-77 A &1.36 &0.90 &1.67 &$6.2^{+4.0}_{-3.5}$ &(11) \\
WASP-173 A &1.39 &1.05 &3.69 &$6.78(2.93)$ &(12) \\
WASP-135  &1.40 &0.98 &1.90 &$0.60^{+1.40}_{-0.35}$ &(13) \\
Qatar-1 &1.42 &0.85 &1.32 &$11.6^{+0.60}_{-4.70}$ &(2) \\
WASP-46 &1.43 &0.83 &1.90 &$9.6^{+3.7}_{-4.2}$ &(2) \\
OGLE-TR-113 &1.43 &0.78 &1.36 &$13.2^{+0.8}_{-2.4}$ &(2) \\
TrES-5 &1.48 &0.89 &1.79 &$7.4(1.9)$ &(2) \\
Kepler-17 &1.49 &1.07 &2.34 &$1.5^{+10.2}_{-1.9}$ &(14) \\
\bottomrule
\end{tabular}
\end{table}

The orbital decay of WASP-12b can be understood if the host star is a subgiant with $\Ms\simeq 1.2 M_\sun$; the linearly excited primary mode (the parent) is then in the wave breaking regime and the expected $\dot{P}_{\rm orb}$ agrees well with the observed value \citep{Weinberg:2017}.  However, \citet{Bailey:2019} find that subgiant models are in greater tension with current observational constraints than main-sequence models with $\Ms\simeq 1.3-1.4 M_\sun$.  Since such a main-sequence star has a convective core, the excited primary mode should be a standing wave, i.e., a $g$-mode, and stable to the parametric instability.  It is then not clear how to explain the observed decay since the linear and nonlinear dissipation of the $g$-mode is too small.

\begin{table}[!t]\centering
\caption{\label{tab:high_mass_data}
Favorable targets for the search for tidal orbital decay for host stars with $\Ms > 1.1 M_\sun$.   We exclude WASP-12 and Kepler-1658 from this list since their orbital decay has been detected (see Section~\ref{sec:wasp12}). The age column includes the uncertainty, which is symmetric about the best fit if enclosed in parentheses.  The fractional uncertainties are generally $\lesssim 0.001\%$ for $\Porb$  and  $\lesssim 10\%$ for $\Ms$ and $\Mp$.\\
{\bf References.}
(1) \citet{Wong:2021}; 
(2) \citet{Bonomo:2017}; 
(3) \citet{Oberst:2017}; 
(4) \citet{Delrez:2016}; 
(5) \citet{Esteves:2015}; 
(6) \citet{Barros:2016}; 
(7) \citet{Alsubai:2019}; 
(8) \citet{Turner:2016}; 
(9) \citet{Deleuil:2014}; 
(10) \citet{West:2016}; 
(11) \citet{deVal-Borro:2016};
(12) \citet{Yee:2023}. 
}
\begin{tabular}{lccccc}\toprule
&$P_{\rm orb}$ &$\Ms$ &$\Mp$ &Age & \\\cmidrule{2-5}
Hostname &(days) &($M_\sun$) &($\Mj$) &(Gyr) &Refs. \\\midrule
TOI-2109 &0.67 &1.45 &5.02 &$1.77(0.88)$ & (1) \\
WASP-103 &0.93 &1.22 &1.455 &$4(1)$ & (2)\\
KELT-16 &0.97 &1.21 &2.75 &$3.1(0.3)$ & (3)\\
OGLE-TR-56 &1.21 &1.23 &3.3 &$3.0^{+2.9}_{-1.4}$ & (2) \\
HAT-P-23 &1.21 &1.13 &1.97 &$4.0(1.0)$ & (2)\\
WASP-121 &1.27 &1.35 &1.183 &$1.5(1.0)$ & (4)\\
CoRoT-14 &1.51 &1.13 &7.42 &$4.2(3.8)$ & (2)\\
Kepler-76 &1.54 &1.20 &2.0 &$3.6^{+3.7}_{-1.3}$ & (5) \\
WASP-114 &1.55 &1.29 &1.769 &$4.3^{+1.4}_{-1.3}$ & (6)\\
Qatar-10 &1.65 &1.16 &0.736 &$3.2(1.9)$ & (7) \\
KELT-14 &1.71 &1.24 &1.284 &$5.11(0.80)$ & (8)\\
Kepler-412 &1.72 &1.17 &0.939 &$5.1(1.7)$ & (9)\\
WASP-76 &1.81 &1.46 &0.92 &$5.3^{+6.1}_{-2.9}$ & (10)\\
HATS-35 &1.82 &1.32 &1.222 &$2.13(0.51)$ & (11) \\
TOI-2803 A &1.96 &1.12 &0.975 &$3.7^{+1.5}_{-1.3}$ & (12)\\
\bottomrule
\end{tabular}
\end{table}

Kepler-1658  has a mass $\Ms\simeq 1.5 M_\sun$ and is more definitively a subgiant star. \citet{Vissapragada:2022} explain that the measured decay is in good agreement with theoretical predictions for inertial wave dissipation \citep{Barker:2020}.  Such dissipation is possible because the tidal frequency $2(\Omega-\Omega_{\rm spin})$ is less than twice the star's spin frequency $\Omega_{\rm spin}$ (this is because of the relatively large $\Porb=3.85\trm{ days}$ and the star's relatively short spin period $P_{\rm spin}= 2\pi/\Omega_{\rm spin} =5.66 \trm{ days}$; WASP-12b orbits too quickly and the host star spins too slowly for the tide to excite inertial waves).   Although the linearly excited gravity wave  breaks as it approaches the stellar center,  we find that the wave luminosity is too small to explain the observed orbital decay (here we are ignoring rotation; the wave luminosity  scales approximately as $\Porb^{-23/3}$ and thus it is much smaller in Kepler-1658 than in a subgiant model of WASP-12).
 
\subsection{Systems likely to be undergoing rapid orbital decay}
\label{sec:promising}

In Tables~\ref{tab:low_mass_data} and \ref{tab:high_mass_data} we list hot Jupiter systems that our calculations suggest are favorable targets for the search for orbital decay, separated into systems with $\Ms \le 1.1 M_\sun$ and $\Ms > 1.1 M_\sun$, respectively.  Given the strong dependence of $\aveT$ on $\Porb$, these tables essentially list the systems with especially small $\Porb$.  However, we exclude short $\Porb$ systems whose host stars are likely to have convective cores (main sequence stars with $\Ms \gtrsim 1.2 M_\sun$) since we find that such stars are  unlikely to be undergoing significant orbital decay; examples of such systems include WASP-18 ($\Porb=0.94\trm{ days}$, $\Ms=1.3 M_\sun$, $\trm{Age}=1.6^{+1.4}_{-0.9}\trm{ Gyr}$; \citealt{Cortes-Zuleta:2020}), HATS-52 ($\Porb=1.37\trm{ days}$, $\Ms=1.1 M_\sun$, $\trm{Age}=1.2^{+1.5}_{-1.1}\trm{ Gyr}$; \citealt{Henning:2018}), and CoRoT-1 ($\Porb=1.51\trm{ days}$, $\Ms=1.2 M_\sun$, $\trm{Age}=1.6\pm0.5\trm{ Gyr}$; \citealt{Bonomo:2017}). That said, the systems listed in Table~\ref{tab:high_mass_data}  have sufficiently large age uncertainties that many are likely still on the main sequence, in which case their $\aveT$ would be large.

For the systems listed in Table~\ref{tab:low_mass_data} with $\Porb \lesssim 1.0\trm{ day}$, our results suggest $\aveT \lesssim 100 \trm{ Myr}$ and $\Q \lesssim 10^6$  (given the uncertainties in the measured  parameters, especially the stellar age, we do not attempt to make system specific predictions). If a system listed in Table~\ref{tab:high_mass_data} has a host star on the subgiant branch, then it too should have $\aveT \lesssim 100 \trm{ Myr}$ and $\Q \lesssim 10^6$.   By Equation~(\ref{eq:Tshift}), we expect systems with such small $\aveT$ to have $\Tshift \gtrsim 10 \trm{ s}$ after $T_{\rm dur}=10\trm{ yr}$.  Thus, the transit timing shifts of some of these systems, many of which are part of existing transit timing measurement campaigns (e.g., \citealt{Patra:2020,  Ivshina:2022, Maciejewski:2022, Mannaday:2022, Rosario:2022,  Harre:2023}), may soon be detectable. 

\section{\bf S\lowercase{ummary and} C\lowercase{onclusions}}
\label{sec:conclusions}

We studied the dissipation of the dynamical tide due to the excitation and damping of weakly nonlinear $g$-modes in hot Jupiter host stars.  This is the  dominant source of tidal dissipation in the great majority of hot Jupiter systems. By integrating the amplitude equations for large networks of nonlinearly interacting $g$-modes consisting of a linearly driven parent exciting a sea of secondary modes (daughters and granddaughters), we calculated the tidal dissipation rate as a function of stellar mass $\Ms$, age, orbital period $\Porb$, and planetary mass $\Mp$.  We expressed our results in terms of the orbital decay time $\aveT$, stellar tidal quality factor $\Q$, and transit time shift $\Tshift$. In order to span the range of observed hot Jupiter systems, our analysis considered 16 stellar models with masses between $\Ms=0.5-2.0 M_\sun$ and ages between the early main sequence and the subgiant branch. For each stellar model we considered orbital periods between $\Porb=0.5-4\trm{ days}$ and planetary masses between $\Mp=0.5-3.0 \Mj$. Our main results are shown in Figures~\ref{fig:tdecay_low_mass}--\ref{fig:tdecay_high_mass} and in Table~\ref{tab:tab_run_data}.

For hosts with $\Ms\lesssim 1.0 M_\sun$, we found that the tidal dissipation rate increases with $\Ms$ and age, and can be significant throughout the main sequence.  The dissipation is sensitive to $\Porb$ and in many cases follows a power-law $\aveT \propto \Porb^\alpha$ with $\alpha\simeq 6-7$. For $\Porb\lesssim 1.0 \trm{ day}$ and $\Mp\gtrsim 0.5\Mj$ we in general found that $\aveT \lesssim 100 \trm{ Myr}$, $\Q \lesssim 10^6$, and $\Tshift \gtrsim 10 (T_{\rm dur}/10\trm{ yr})^2 \trm{ s}$, where $T_{\rm dur}$ is the duration of a system's  observation time.

For hosts with $\Ms \gtrsim 1.2 M_\sun$, we found that the tidal dissipation rate is negligible on the main sequence but becomes highly significant as these stars ascend the subgiant branch (owing primarily to the respective presence and absence of a convective core).  Even at $\Porb \approx 3\trm{ days}$, the dissipation rate on the subgiant branch is rapid enough to produce a detectable transit time shift ($\Tshift \gtrsim 10\trm{ s}$) within 10 years. For $\Porb\approx 1.0\trm{ day}$, a detectable transit time shift takes only $\approx 1\trm{ yr}$, which is $\sim 100$ times faster than for main sequence stars with $\Ms\lesssim 1.0 M_\sun$ (see Figure~\ref{fig:Tshift}).

We compared our results to known hot Jupiters and identified a number of systems that could be undergoing rapid orbital decay (see Tables~\ref{tab:low_mass_data} and \ref{tab:high_mass_data}). However, since our results are sensitive to the age of the star (especially for $\Ms \gtrsim 1.2 M_\sun$), which is usually the least well-measured parameter of the hosts, there can be considerable uncertainty in the expected $\aveT$ and $\Tshift$ of individual systems.

Our analysis assumed that the excited modes are all global standing waves.  We found that this is a good approximation for the linearly excited parent, except perhaps in our most evolved subgiant models.  In those models, the parent's radiative damping can exceed its group travel time through the star or its displacement amplitude $\xi_r$ can approach the wave breaking limit ($k_r \xi_r\gtrsim 1$) near the stellar center. If either is true, the parent should instead be treated as a traveling wave since it dissipates all of its energy and angular momentum before it can reflect near the stellar center and form a standing wave.  Several studies have considered the tidal dissipation rate in hot Jupiter systems under the assumption that the parent is a traveling wave and thus maximally dissipative  \citep{Barker:2010, Barker:2011b, Chernov:2017, Weinberg:2017, Sun:2018, Lazovik:2021}.

Treating all the daughters and granddaughters as standing waves may not always be a good approximation, however, even for some of our main sequence models. This is because their wavelengths can be much shorter than the parent that excites them.  They therefore have higher radiative damping rates and at a given mode energy are closer to the wave breaking limit than the parent.  This suggests that the interaction between the parent and the secondary modes may in some cases involve a standing wave (the parent) nonlinearly exciting traveling waves (the daughters, granddaughters, etc.).  The computational methods for studying weakly nonlinear interactions between standing waves and traveling waves have not been developed, as far as we know, and is left for future work. Since the secondary modes dissipate their energy by radiative damping far from the stellar center where they are nonlinearly excited,  it is possible the total dissipation will be insensitive to whether they are treated as standing waves or traveling waves.

Another caveat of our calculation is that we do not account for possible changes to the stellar structure due to the transfer of energy and angular momentum from the modes to the background star. This type of interaction was investigated recently by \citet{Guo:2023}, who performed 2D hydrodynamical simulations of tidally excited nonlinear gravity waves in the cores of solar-type stars. They found that linear damping of the waves gradually spins up the core and that subsequent incoming (parent) waves are absorbed in an expanding critical layer.  Importantly, this process was found to occur even when the parents are below the wave breaking threshold, suggesting that such parents are effectively in the traveling wave regime and thus maximally dissipative.  Due to computational limitations, they say that the secondary waves generated by the parametric instability, which are the focus of our analysis,  would be very difficult to observe in their simulations.  Properly resolving such secondary waves might be important, however.  For example, they could  propagate outwards from where they are excited and transfer their energy and angular momentum at larger radii rather than locally.  Their dissipation might also modify the stratification of the central region, moving the parent's inner turning point outwards and causing it to reflect at larger radii. Such effects can potentially impact the formation of a critical layer and the subsequent absorption of the incoming parent wave.

\begin{acknowledgments}
This work was supported by NSF grant No. AST-2054353. 
\end{acknowledgments}

\software{\texttt{MESA } \citep[][\url{http://mesa.sourceforge.net}]{Paxton:11, Paxton:13, Paxton:15, Paxton:18,Paxton:19, Jermyn:2022},
\texttt{GYRE } \citep[][\url{https://bitbucket.org/rhdtownsend/gyre/wiki/Home}]{Townsend:13, Townsend:18},  \texttt{SUNDIAL }\citep[\url{https://computing.llnl.gov/projects/sundials}]{Hindmarsh:05}.
}

\facility {Exoplanet Archive} This research has made use of the NASA Exoplanet Archive, which is operated by the California Institute of Technology, under contract with the National Aeronautics and Space Administration under the Exoplanet Exploration Program.

\appendix

\section{Parameters of the \texttt{MESA} inlists}
\label{sec:mesa_inlist}

The stellar models are built using version 15140 of \texttt{MESA} \citep{Paxton:11,Paxton:13,Paxton:15,Paxton:18,Paxton:19,Jermyn:2022}. The key parameters of the inlist files we use are provided below. We change \verb|initial_mass| to build  different mass models and we change \verb|max_center_cell_dq|, \verb|R_function_weight|, and \verb|R_function_param| if finer grid resolution is needed in the stellar center (e.g., to calculate $\kappa_{abc}$).

\begin{verbatim}
&star_job
    create_pre_main_sequence_model = .true.
/ ! end of star_job namelist

&controls
 initial_mass = 0.5
 initial_z = 0.02d0
 use_Ledoux_criterion = .true.
 MLT_option = `Henyey'
 max_center_cell_dq = 1d-10
 R_function_weight = 100
 R_function_param = 1d-8     
 use_dedt_form_of_energy_eqn = .true.
 use_gold_tolerances = .true.
 mesh_delta_coeff = 0.2
 when_to_stop_rtol = 1d-6
 when_to_stop_atol = 1d-6
/ ! end of controls namelist

\end{verbatim}

\section{Table of orbital decay results from the mode network integrations}
\label{sec:table_of_results}

In Table~\ref{tab:tab_run_data} we list the values of $\aveT$, $\Q$, and $\Tshift$ from each of our mode network integrations (see also Figures~\ref{fig:tdecay_low_mass}, \ref{fig:Tshift}, and \ref{fig:tdecay_high_mass}).

     \begin{longtable*}{c | c c c c}
       \caption{Orbital decay time $\aveT$, stellar tidal quality factor $\Q$, and transit time shift $\Tshift$ from the mode network integrations.  Each column is for a particular stellar model ($\Ms$ and Age) and each triplet of rows is, from top to bottom, for an orbital period $\Porb=0.5, 1.0, 2.0, 3.0, 4.0 \trm{ days}$. For a given $\Porb$, the first row gives $\log_{10} (\aveT / \trm{yr})$, the second row gives $\log_{10} \Q$, and the third row gives $\log_{10} (\Tshift / \trm{s})$ assuming $T_{\rm dur}=10\trm{ yr}$.  The four comma-separated entries in each row are, from left to right, for planet mass $\Mp=0.5,1.0,2.0,3.0 \Mj$. The top of the table illustrates the format.
      \label{tab:tab_run_data}}\\
      \toprule
      $\Porb$ & $\Ms$, Age & $\Ms$, Age & $\Ms$, Age & $\Ms$, Age \\
      (day) & ($M_\sun$, Gyr) & ($M_\sun$, Gyr) & ($M_\sun$, Gyr) & ($M_\sun$, Gyr) 
      \\ [1ex]
      \hline
      \hline
      $-$ &$\log_{10} (\aveT / \trm{yr})$: $0.5,1.0,2.0,3.0 \Mj$&  $-$ & $-$ & $-$ \\
      $-$ &   $\log_{10} \Q$: $0.5,1.0,2.0,3.0 \Mj$&  $-$ & $-$ & $-$ \\
      $-$ &$\log_{10} (\Tshift / \trm{s})$: $0.5,1.0,2.0,3.0 \Mj$&  $-$ & $-$ & $-$ \\
      \hline
      \hline
      \noalign{\vskip 1mm} 
      $-$ & $0.5 M_\sun$, 1.07 Gyr & $0.5 M_\sun$, 4.97 Gyr & $0.5 M_\sun$, 8.97 Gyr & $-$\\
      [1ex]
      \hline
      0.5 &  $16.1, 15.8, 15.5, 15.4 $  &  $ 8.5,  7.5,  6.9,  6.4 $ &  $  8.6,  8.3,  8.1,  7.1  $  & $-$   \\
      $-$  & $ 14.1, 14.1, 14.1, 14.2$   & $  6.6,  5.9,  5.6,  5.3$  & $   6.5,  6.6,  6.6,  5.8 $   &$-$   \\
      $-$  & $ -6.8, -6.5, -6.2, -6.0$   & $  0.8,  1.9,  2.5,  2.9$  & $   0.8,  1.1,  1.3,  2.2 $   &$-$   \\[1.5ex] 
      1.0 &  $16.9, 16.6, 16.3, 15.7 $  &  $13.0,  9.9,  9.1,  8.7 $ &  $ 10.6, 10.0,  9.4,  9.0  $  &$-$   \\
      $-$  & $ 13.5, 13.5, 13.5, 13.0$   & $  9.7,  6.9,  6.4,  6.2$  & $   7.3,  7.0,  6.7,  6.4 $   &$-$   \\
      $-$  & $ -7.5, -7.2, -7.0, -6.3$   & $ -3.6, -0.5,  0.3,  0.7$  & $  -1.2, -0.6, -0.0,  0.4 $   &$-$   \\[1.5ex] 
       2.0 & $18.2, 17.8, 17.3, 17.1 $  &  $16.7, 12.9, 11.6, 11.2 $ &  $ 14.7, 12.7, 11.7, 11.3  $  &$-$   \\
      $-$  & $ 13.5, 13.5, 13.3, 13.2$   & $ 12.0,  8.6,  7.5,  7.4$  & $  10.0,  8.4,  7.7,  7.4 $   &$-$   \\
      $-$  & $ -8.8, -8.5, -8.0, -7.8$   & $ -7.3, -3.5, -2.2, -1.9$  & $  -5.3, -3.4, -2.3, -1.9 $   &$-$   \\[1.5ex] 
       3.0 & $18.4, 18.1, 17.8, 17.6 $  &  $17.3, 15.4, 14.1, 13.1 $ &  $ 16.2, 15.1, 13.1, 12.1  $  &$-$   \\
      $-$  & $ 12.9, 12.9, 12.9, 12.9$   & $ 11.9, 10.3,  9.3,  8.5$  & $  10.8, 10.0,  8.3,  7.4 $   &$-$   \\ 
      $-$  & $ -9.0, -8.7, -8.4, -8.2$   & $ -7.9, -6.0, -4.7, -3.7$  & $  -6.8, -5.7, -3.7, -2.7 $   &$-$   \\[1.5ex] 
       4.0 & $18.8, 18.5, 18.2, 18.0 $  &  $17.9, 17.1, 16.6, 15.4 $ &  $ 18.3, 16.4, 14.8, 13.5  $  &$-$   \\
      $-$  & $ 12.8, 12.8, 12.8, 12.8$   & $ 12.0, 11.4, 11.3, 10.3$  & $  12.4, 10.7,  9.5,  8.4 $   &$-$   \\ 
      $-$  & $ -9.4, -9.1, -8.8, -8.6$   & $ -8.5, -7.7, -7.3, -6.1$  & $  -9.0, -7.0, -5.5, -4.2 $   &$-$   \\ 
       [1.5ex] 
      \hline
      \hline \noalign{\vskip 1mm} 
      & $0.8 M_\sun$, 1.07 Gyr & $0.8 M_\sun$, 4.87 Gyr & $0.8 M_\sun$, 8.97 Gyr & $-$ \\
      [1ex]
      \hline
      0.5 &  $ 8.4,  6.6,  6.0,  5.7  $ &   $ 6.6,  6.2,  6.0,  5.9  $ &  $ 6.3,  6.2,  6.1,  6.0  $ &$-$  \\
     $-$  &  $  6.8,  5.3,  5.0,  4.9 $  &  $  5.0,  4.9,  5.0,  5.1 $  & $  4.9,  5.0,  5.2,  5.3 $  &$-$  \\
     $-$  &  $  1.0,  2.8,  3.4,  3.7 $  &  $  2.8,  3.2,  3.4,  3.5 $  & $  3.0,  3.2,  3.3,  3.4 $  &$-$  \\[1.5ex] 
     1.0 &   $11.7,  8.7,  8.3,  8.2  $ &   $ 8.9,  8.5,  8.0,  7.8  $ &  $ 8.5,  8.2,  8.0,  7.9  $ &$-$  \\
     $-$  &  $  8.9,  6.2,  6.1,  6.1 $  &  $  6.1,  6.0,  5.8,  5.8 $  & $  5.8,  5.7,  5.9,  6.0 $  &$-$  \\
     $-$  &  $ -2.3,  0.7,  1.1,  1.2 $  &  $  0.5,  0.9,  1.4,  1.6 $  & $  0.9,  1.2,  1.4,  1.4 $  &$-$  \\[1.5ex] 
      2.0 &  $16.1, 15.3, 10.8, 10.1  $ &   $11.2, 10.7, 10.2,  9.9  $ &  $11.0, 10.5, 10.0,  9.9  $ &$-$  \\
     $-$  &  $ 11.9, 11.5,  7.2,  6.8 $  &  $  7.1,  6.9,  6.7,  6.5 $  & $  6.9,  6.8,  6.6,  6.6 $  &$-$  \\
     $-$  &  $ -6.7, -6.0, -1.4, -0.8 $  &  $ -1.9, -1.3, -0.8, -0.5 $  & $ -1.6, -1.2, -0.6, -0.5 $  &$-$  \\[1.5ex] 
      3.0 &  $16.6, 16.3, 14.2, 13.6  $ &   $13.3, 12.4, 11.5, 11.2  $ &  $12.1, 11.7, 11.2, 11.1  $ &$-$  \\
     $-$  &  $ 11.7, 11.7,  9.8,  9.5 $  &  $  8.4,  7.9,  7.2,  7.1 $  & $  7.3,  7.2,  7.0,  7.1 $  &$-$  \\ 
     $-$  &  $ -7.2, -6.9, -4.8, -4.2 $  &  $ -3.9, -3.1, -2.1, -1.9 $  & $ -2.8, -2.3, -1.9, -1.7 $  &$-$  \\[1.5ex] 
      4.0 &  $17.0, 16.3, 15.4, 14.9  $ &   $15.0, 14.2, 12.2, 12.3  $ &  $13.5, 12.9, 12.4, 12.1  $ &$-$  \\
     $-$  &  $ 11.5, 11.1, 10.5, 10.2 $  &  $  9.5,  9.0,  7.4,  7.6 $  & $  8.1,  7.8,  7.7,  7.5 $  &$-$  \\ 
     $-$  &  $ -7.6, -6.9, -6.0, -5.5 $  &  $ -5.6, -4.8, -2.9, -2.9 $  & $ -4.1, -3.5, -3.0, -2.7 $  &$-$  \\ 
       [1.5ex] 
      \hline
      \hline \noalign{\vskip 1mm} 
      & $1.0 M_\sun$, 1.03 Gyr & $1.0 M_\sun$, 4.60 Gyr & $1.0 M_\sun$, 8.72 Gyr  & $1.0 M_\sun$, 11.0 Gyr \\
      [1ex]
      \hline
      0.5 &  $  7.2,  6.4,  5.8,  5.6  $ &  $  6.3,  6.1,  6.0,  5.9  $ &   $ 6.0,  5.9,  5.7,  5.7  $ &  $   5.7,  5.5,  5.4,  5.4  $ \\
      $-$  & $   5.8,  5.3,  5.0,  5.0 $  & $   5.2,  5.3,  5.5,  5.6 $  &  $  5.3,  5.5,  5.6,  5.7 $  & $    5.5,  5.7,  5.8,  6.0 $  \\ 
      $-$  & $   2.2,  3.0,  3.6,  3.8 $  & $   3.1,  3.2,  3.4,  3.5 $  &  $  3.4,  3.5,  3.6,  3.7 $  & $    3.7,  3.8,  4.0,  4.0 $  \\[1.5ex] 
      1.0 &  $  9.4,  8.4,  8.2,  7.9  $ &  $  8.1,  8.0,  7.8,  7.6  $ &   $ 7.8,  7.7,  7.6,  7.5  $ &  $   7.1,  7.0,  6.8,  6.7  $ \\
      $-$  & $   6.8,  6.1,  6.1,  6.0 $  & $   5.7,  5.8,  5.7,  6.0 $  &  $  5.8,  5.9,  6.1,  6.2 $  & $    5.7,  5.8,  5.9,  6.0 $  \\ 
      $-$  & $  -0.0,  0.9,  1.2,  1.5 $  & $   1.3,  1.4,  1.5,  1.8 $  &  $  1.6,  1.7,  1.8,  1.9 $  & $    2.2,  2.4,  2.6,  2.7 $  \\[1.5ex] 
      2.0 &  $ 14.2, 12.6,  9.9,  9.9  $ &  $ 10.5,  9.9,  9.7,  9.5  $ &   $ 9.8,  9.6,  9.5,  9.4  $ &  $   9.5,  9.4,  9.3,  9.2  $ \\
      $-$  & $  10.3,  9.0,  6.6,  6.7 $  & $   6.7,  6.5,  6.5,  6.6 $  &  $  6.4,  6.6,  6.8,  6.8 $  & $    6.8,  6.9,  7.1,  7.2 $  \\ 
      $-$  & $  -4.8, -3.2, -0.6, -0.5 $  & $  -1.1, -0.6, -0.3, -0.2 $  &  $ -0.4, -0.2, -0.1, -0.0 $  & $   -0.2, -0.0,  0.1,  0.2 $  \\[1.5ex] 
      3.0 &  $ 15.0, 14.5, 13.1, 12.2  $ &  $ 11.8, 11.3, 11.2, 11.0  $ &   $ 1.0, 11.0, 10.8, 10.7  $ &  $  11.4, 10.6, 10.4, 10.4  $ \\
      $-$  & $  10.3, 10.1,  9.1,  8.3 $  & $   7.3,  7.1,  7.3,  7.3 $  &  $  6.9,  7.2,  7.3,  7.4 $  & $    7.8,  7.4,  7.5,  7.6 $  \\ 
      $-$  & $  -5.6, -5.1, -3.8, -2.8 $  & $  -2.4, -1.9, -1.8, -1.7 $  &  $ -1.6, -1.6, -1.4, -1.4 $  & $   -2.0, -1.3, -1.1, -1.0 $  \\[1.5ex] 
      4.0 &  $ 16.4, 16.1, 14.3, 14.0  $ &  $ 12.7, 12.4, 11.8, 11.8  $ &   $ 1.7, 11.6, 11.4, 11.3  $ &  $  11.6, 11.7, 11.6, 11.5  $ \\
      $-$  & $  11.2, 11.2,  9.7,  9.5 $  & $   7.7,  7.7,  7.3,  7.5 $  &  $  7.0,  7.2,  7.4,  7.5 $  & $    7.5,  7.9,  8.1,  8.2 $  \\ 
      $-$  & $  -7.0, -6.7, -4.9, -4.6 $  & $  -3.4, -3.0, -2.4, -2.4 $  &  $ -2.3, -2.2, -2.0, -1.9 $  & $   -2.2, -2.3, -2.2, -2.2 $  \\ 
      [1.5ex] 
      \hline
      \hline \noalign{\vskip 1mm} 
      &$-$ &$-$ & $1.2 M_\sun$, 4.00 Gyr & $1.2 M_\sun$, 6.00 Gyr  \\
       [1ex]
       \hline
      0.5 & $-$ &$-$  &  $ 6.1,  6.0,  5.8,  5.8  $  &   $ 3.4,  3.3,  3.2,  3.2  $ \\
      $-$  & $-$ &$-$  & $  5.7,  5.8,  6.0,  6.1 $   &  $  5.1,  5.3,  5.5,  5.6 $  \\ 
      $-$  & $-$ &$-$  & $  3.3,  3.4,  3.5,  3.6 $   &  $  6.0,  6.1,  6.1,  6.2 $  \\[1.5ex] 
      1.0 & $-$ &$-$  &  $ 8.0,  7.8,  7.7,  7.6  $  &   $ 5.5,  5.4,  5.2,  5.2  $ \\
      $-$  & $-$ &$-$  & $  6.2,  6.4,  6.5,  6.6 $   &  $  5.9,  6.0,  6.2,  6.3 $  \\ 
      $-$  & $-$ &$-$  & $  1.4,  1.6,  1.7,  1.8 $   &  $  3.9,  4.0,  4.1,  4.2 $  \\[1.5ex] 
      2.0 & $-$ &$-$  &  $10.0, 10.0,  9.8,  9.7  $  &   $ 7.6,  7.6,  7.6,  7.6  $\\
      $-$  & $-$ &$-$  & $  7.0,  7.2,  7.4,  7.4 $   &  $  6.6,  6.9,  7.2,  7.4 $  \\ 
      $-$  & $-$ &$-$  & $ -0.7, -0.6, -0.4, -0.3 $   &  $  1.8,  1.8,  1.8,  1.8 $  \\[1.5ex] 
      3.0 & $-$ &$-$  &  $11.1, 11.1, 11.1, 11.1  $  &   $ 8.1,  8.1,  8.1,  8.1  $ \\
      $-$  & $-$ &$-$  & $  7.3,  7.6,  7.9,  8.0 $   &  $  6.4,  6.7,  7.0,  7.2 $  \\ 
      $-$  & $-$ &$-$  & $ -1.7, -1.7, -1.7, -1.7 $   &  $  1.3,  1.3,  1.3,  1.3 $  \\[1.5ex] 
      4.0 & $-$ &$-$  &  $11.9, 11.5, 11.4, 11.4  $  &   $ 8.8,  8.4,  8.4,  8.4  $ \\
      $-$  & $-$ &$-$  & $  7.6,  7.5,  7.6,  7.8 $   &  $  6.6,  6.5,  6.8,  7.0 $  \\ 
      $-$  & $-$ &$-$  & $ -2.5, -2.1, -2.0, -2.0 $   &  $  0.6,  0.9,  1.0,  0.9 $  \\ 
       [1.5ex] 
       \hline
       \hline \noalign{\vskip 1mm} 
       &$-$ &$-$  &  $1.5 M_\sun$, 2.03 Gyr & $1.5 M_\sun$, 2.5 Gyr \\
        [1ex]
        \hline
       0.5 &$-$ &$-$  &  $  7.5,  7.3,  7.2,  7.2  $ &  $  3.4,  3.3,  3.2,  3.1   $ \\
       $-$  &$-$ &$-$  & $   7.6,  7.8,  8.0,  8.1 $  & $   4.5,  4.6,  4.8,  4.9  $  \\ 
       $-$  &$-$ &$-$  & $   1.9,  2.0,  2.1,  2.2 $  & $   5.9,  6.1,  6.2,  6.3  $  \\[1.5ex] 
       1.0 &$-$ &$-$  &  $  9.5,  9.2,  9.0,  8.9  $ &  $  5.7,  5.6,  5.5,  5.4   $ \\
       $-$  &$-$ &$-$  & $   8.4,  8.4,  8.4,  8.5 $  & $   5.5,  5.6,  5.8,  5.9  $  \\ 
       $-$  &$-$ &$-$  & $  -0.2,  0.1,  0.4,  0.5 $  & $   3.6,  3.8,  3.9,  4.0  $  \\[1.5ex] 
       2.0 &$-$ &$-$  &  $ 11.8, 11.2, 11.2, 11.2  $ &  $  7.8,  7.8,  7.7,  7.7   $ \\
       $-$  &$-$ &$-$  & $   9.3,  9.1,  9.3,  9.5 $  & $   6.2,  6.5,  6.8,  6.9  $  \\ 
       $-$  &$-$ &$-$  & $  -2.4, -1.9, -1.8, -1.8 $  & $   1.6,  1.6,  1.6,  1.7  $  \\[1.5ex] 
       3.0 &$-$ &$-$  &  $ 13.0, 12.7, 12.1, 12.0  $ &  $  8.2,  8.3,  8.3,  8.3   $ \\
       $-$  &$-$ &$-$  & $   9.7,  9.8,  9.5,  9.6 $  & $   5.9,  6.3,  6.6,  6.8  $  \\ 
       $-$  &$-$ &$-$  & $  -3.6, -3.4, -2.7, -2.6 $  & $   1.1,  1.1,  1.0,  1.0  $  \\[1.5ex] 
       4.0 &$-$ &$-$  &  $ 14.5, 13.3, 13.0, 12.8  $ &  $   9.2 , 8.8,  8.6,  8.6  $  \\
       $-$  &$-$ &$-$  & $  10.7,  9.8,  9.8,  9.8 $  & $   6.3 , 6.2,  6.3,  6.5 $   \\ 
       $-$  &$-$ &$-$  & $  -5.1, -3.9, -3.6, -3.4 $  & $    0.2, 0.6,  0.8,  0.8 $   \\ 
        [1.5ex] 
        \hline
        \hline \noalign{\vskip 1mm} 
        &$-$ &$-$  &  $2.0 M_\sun$, 0.70 Gyr & $2.0 M_\sun$, 0.92 Gyr \\
         [1ex]
         \hline
        0.5 &$-$ &$-$ &  $  9.2,  8.5,  7.9,  7.7  $ &  $ 3.4,  3.2,  3.1,  3.0  $ \\
        $-$  &$-$ &$-$ & $   9.1,  8.7,  8.5,  8.4 $  & $  5.0,  5.2,  5.4,  5.5 $  \\ 
        $-$  &$-$ &$-$ & $   0.2,  0.9,  1.4,  1.7 $  & $  6.0,  6.1,  6.3,  6.4 $  \\[1.5ex] 
        1.0 &$-$ &$-$ &  $ 12.2, 11.8, 10.5,  9.8  $ &  $ 5.4,  5.2,  5.1,  5.0  $ \\
        $-$  &$-$ &$-$ & $  10.8, 10.7,  9.7,  9.2 $  & $  5.7,  5.9,  6.1,  6.2 $  \\ 
        $-$  &$-$ &$-$ & $  -2.8, -2.5, -1.2, -0.4 $  & $  4.0,  4.1,  4.3,  4.4 $  \\[1.5ex] 
        2.0 &$-$ &$-$ &  $ 14.2, 13.2, 12.8, 12.3  $ &  $ 7.4,  7.5,  7.0,  6.9  $ \\
        $-$  &$-$ &$-$ & $  11.5, 10.9, 10.7, 10.4 $  & $  6.4,  6.9,  6.6,  6.7 $  \\ 
        $-$  &$-$ &$-$ & $  -4.9, -3.9, -3.5, -2.9 $  & $  2.0,  1.8,  2.4,  2.5 $  \\[1.5ex] 
        3.0 &$-$ &$-$ &  $ 14.8, 14.5, 14.2, 14.0  $ &  $ 8.1,  8.1,  8.1,  8.1  $ \\
        $-$  &$-$ &$-$ & $  11.4, 11.4, 11.4, 11.4 $  & $  6.4,  6.7,  7.0,  7.1 $  \\ 
        $-$  &$-$ &$-$ & $  -5.4, -5.1, -4.8, -4.7 $  & $  1.3,  1.3,  1.3,  1.3 $  \\[1.5ex] 
        4.0 &$-$ &$-$ &  $ 15.2, 14.9, 14.6, 14.5  $ &  $ 8.5,  8.4,  8.4,  8.4  $ \\
        $-$  &$-$ &$-$ & $  11.2, 11.2, 11.2, 11.2 $  & $  6.2,  6.4,  6.7,  7.0 $  \\ 
        $-$  &$-$ &$-$ & $  -5.9, -5.6, -5.3, -5.1 $  & $  0.9,  1.0,  1.0,  0.9 $  \\ 
      \bottomrule
      \end{longtable*}

\bibliography{refs}{}

\begin{thebibliography}{}
\expandafter\ifx\csname natexlab\endcsname\relax\def\natexlab#1{#1}\fi
\providecommand{\url}[1]{\href{#1}{#1}}
\providecommand{\dodoi}[1]{doi:~\href{http://doi.org/#1}{\nolinkurl{#1}}}
\providecommand{\doeprint}[1]{\href{http://ascl.net/#1}{\nolinkurl{http://ascl.net/#1}}}
\providecommand{\doarXiv}[1]{\href{https://arxiv.org/abs/#1}{\nolinkurl{https://arxiv.org/abs/#1}}}

\bibitem[{{Aerts} {et~al.}(2010){Aerts}, {Christensen-Dalsgaard}, \&
  {Kurtz}}]{Aerts:2010}
{Aerts}, C., {Christensen-Dalsgaard}, J., \& {Kurtz}, D. 2010.
\newblock \url{http://dx.doi.org/10.1007/978-1-4020-5803-5}

\bibitem[{{Alsubai} {et~al.}(2019){Alsubai}, {Tsvetanov}, {Pyrzas}, {Latham},
  {Bieryla}, {Eastman}, {Mislis}, {Esquerdo}, {Southworth}, {Mancini},
  {Esamdin}, {Liu}, {Ma}, {Bretton}, {Pall{\'e}}, {Murgas}, {Vilchez},
  {Parviainien}, {Monta{\~n}es-Rodriguez}, {Narita}, {Fukui}, {Kusakabe},
  {Tamura}, {Barkaoui}, {Pozuelos}, {Gillon}, {Jehin}, {Benkhaldoun}, \&
  {Daassou}}]{Alsubai:2019}
{Alsubai}, K., {Tsvetanov}, Z.~I., {Pyrzas}, S., {et~al.} 2019, The
  Astronomical Journal, 157, 224, \dodoi{10.3847/1538-3881/ab19bc}

\bibitem[{{Avallone} {et~al.}(2022){Avallone}, {Tayar}, {van Saders}, {Berger},
  {Claytor}, {Beaton}, {Teske}, {Godoy-Rivera}, \& {Pan}}]{Avallone:2022}
{Avallone}, E.~A., {Tayar}, J.~N., {van Saders}, J.~L., {et~al.} 2022, The
  Astrophysical Journal, 930, 7, \dodoi{10.3847/1538-4357/ac60a1}

\bibitem[{{Bailey} \& {Goodman}(2019)}]{Bailey:2019}
{Bailey}, A., \& {Goodman}, J. 2019, Monthly Notices of the Royal Astronomical
  Society, 482, 1872, \dodoi{10.1093/mnras/sty2805}

\bibitem[{Barker(2011)}]{Barker:2011b}
Barker, A.~J. 2011, Monthly Notices of the Royal Astronomical Society, 414,
  1365, \dodoi{10.1111/j.1365-2966.2011.18468.x}

\bibitem[{{Barker}(2020)}]{Barker:2020}
{Barker}, A.~J. 2020, Monthly Notices of the Royal Astronomical Society, 498,
  2270, \dodoi{10.1093/mnras/staa2405}

\bibitem[{{Barker} \& {Lithwick}(2014)}]{Barker:2014}
{Barker}, A.~J., \& {Lithwick}, Y. 2014, mnras, 437, 305,
  \dodoi{10.1093/mnras/stt1884}

\bibitem[{{Barker} \& {Ogilvie}(2010)}]{Barker:2010}
{Barker}, A.~J., \& {Ogilvie}, G.~I. 2010, mnras, 404, 1849,
  \dodoi{10.1111/j.1365-2966.2010.16400.x}

\bibitem[{{Barker} \& {Ogilvie}(2011)}]{Barker:2011a}
---. 2011, mnras, 417, 745, \dodoi{10.1111/j.1365-2966.2011.19322.x}

\bibitem[{{Barros} {et~al.}(2016){Barros}, {Brown}, {H{\'e}brard}, {G{\'o}mez
  Maqueo Chew}, {Anderson}, {Boumis}, {Delrez}, {Hay}, {Lam}, {Llama}, {Lendl},
  {McCormac}, {Skiff}, {Smalley}, {Turner}, {Vanhuysse}, {Armstrong}, {Boisse},
  {Bouchy}, {Collier Cameron}, {Faedi}, {Gillon}, {Hellier}, {Jehin}, {Liakos},
  {Meaburn}, {Osborn}, {Pepe}, {Plauchu-Frayn}, {Pollacco}, {Queloz}, {Rey},
  {Spake}, {S{\'e}gransan}, {Triaud}, {Udry}, {Walker}, {Watson}, {West}, \&
  {Wheatley}}]{Barros:2016}
{Barros}, S.~C.~C., {Brown}, D.~J.~A., {H{\'e}brard}, G., {et~al.} 2016,
  Astronomy and Astrophysics, 593, A113, \dodoi{10.1051/0004-6361/201526517}

\bibitem[{Birkby {et~al.}(2014)Birkby, Cappetta, Cruz, Koppenhoefer, Ivanyuk,
  Mustill, Hodgkin, Pinfield, Sip{\H o}cz, Kov{\'a}cs, Saglia, Pavlenko,
  Barrado, Bayo, Campbell, Catalan, Fossati, G{\'a}lvez-Ortiz, Kenworthy,
  Lillo-Box, Mart{\'\i}n, Mislis, de~Mooij, Nefs, Snellen, Stoev, Zendejas,
  Burgo, Barnes, Goulding, Haswell, Kuznetsov, Lodieu, Murgas, Palle, Solano,
  Steele, \& Tata}]{Birkby:2014}
Birkby, J.~L., Cappetta, M., Cruz, P., {et~al.} 2014, Monthly Notices of the
  Royal Astronomical Society, 440, 1470, \dodoi{10.1093/mnras/stu343}

\bibitem[{{Bonomo} {et~al.}(2017){Bonomo}, {Desidera}, {Benatti}, {Borsa},
  {Crespi}, {Damasso}, {Lanza}, {Sozzetti}, {Lodato}, {Marzari}, {Boccato},
  {Claudi}, {Cosentino}, {Covino}, {Gratton}, {Maggio}, {Micela}, {Molinari},
  {Pagano}, {Piotto}, {Poretti}, {Smareglia}, {Affer}, {Biazzo}, {Bignamini},
  {Esposito}, {Giacobbe}, {H{\'e}brard}, {Malavolta}, {Maldonado}, {Mancini},
  {Martinez Fiorenzano}, {Masiero}, {Nascimbeni}, {Pedani}, {Rainer}, \&
  {Scandariato}}]{Bonomo:2017}
{Bonomo}, A.~S., {Desidera}, S., {Benatti}, S., {et~al.} 2017, Astronomy and
  Astrophysics, 602, A107, \dodoi{10.1051/0004-6361/201629882}

\bibitem[{{Bouma} {et~al.}(2019){Bouma}, {Winn}, {Baxter}, {Bhatti}, {Dai},
  {Daylan}, {D{\'e}sert}, {Hill}, {Kane}, {Stassun}, {Villasenor}, {Ricker},
  {Vanderspek}, {Latham}, {Seager}, {Jenkins}, {Berta-Thompson}, {Col{\'o}n},
  {Fausnaugh}, {Glidden}, {Guerrero}, {Rodriguez}, {Twicken}, \&
  {Wohler}}]{Bouma:2019}
{Bouma}, L.~G., {Winn}, J.~N., {Baxter}, C., {et~al.} 2019, The Astronomical
  Journal, 157, 217, \dodoi{10.3847/1538-3881/ab189f}

\bibitem[{{Brown} {et~al.}(2011){Brown}, {Latham}, {Everett}, \&
  {Esquerdo}}]{Brown:2011}
{Brown}, T.~M., {Latham}, D.~W., {Everett}, M.~E., \& {Esquerdo}, G.~A. 2011,
  The Astronomical Journal, 142, 112, \dodoi{10.1088/0004-6256/142/4/112}

\bibitem[{{Burkart} {et~al.}(2013){Burkart}, {Quataert}, {Arras}, \&
  {Weinberg}}]{Burkart:13}
{Burkart}, J., {Quataert}, E., {Arras}, P., \& {Weinberg}, N.~N. 2013, \mnras,
  433, 332, \dodoi{10.1093/mnras/stt726}

\bibitem[{{Chernov} {et~al.}(2017){Chernov}, {Ivanov}, \&
  {Papaloizou}}]{Chernov:2017}
{Chernov}, S.~V., {Ivanov}, P.~B., \& {Papaloizou}, J.~C.~B. 2017, Monthly
  Notices of the Royal Astronomical Society, 470, 2054,
  \dodoi{10.1093/mnras/stx1234}

\bibitem[{{Cort{\'e}s-Zuleta} {et~al.}(2020){Cort{\'e}s-Zuleta}, {Rojo},
  {Wang}, {Hinse}, {Hoyer}, {Sanhueza}, {Correa-Amaro}, \&
  {Albornoz}}]{Cortes-Zuleta:2020}
{Cort{\'e}s-Zuleta}, P., {Rojo}, P., {Wang}, S., {et~al.} 2020, Astronomy and
  Astrophysics, 636, A98, \dodoi{10.1051/0004-6361/201936279}

\bibitem[{{Dai} {et~al.}(2017){Dai}, {Winn}, {Yu}, \& {Albrecht}}]{Dai:2017}
{Dai}, F., {Winn}, J.~N., {Yu}, L., \& {Albrecht}, S. 2017, The Astronomical
  Journal, 153, 40, \dodoi{10.3847/1538-3881/153/1/40}

\bibitem[{De {et~al.}(2023)De, MacLeod, Karambelkar, Jencson, Chakrabarty,
  Conroy, Dekany, Eilers, Graham, Hillenbrand, Kara, Kasliwal, Kulkarni, Lau,
  Loeb, Masci, Medford, Meisner, Patel, Quiroga-Nu{\~n}ez, Riddle, Rusholme,
  Simcoe, Sjouwerman, Teague, \& Vanderburg}]{De:2023}
De, K., MacLeod, M., Karambelkar, V., {et~al.} 2023, Nature, 617, 55,
  \dodoi{10.1038/s41586-023-05842-x}

\bibitem[{{de Val-Borro} {et~al.}(2016){de Val-Borro}, {Bakos}, {Brahm},
  {Hartman}, {Espinoza}, {Penev}, {Ciceri}, {Jord{\'a}n}, {Bhatti}, {Csubry},
  {Bayliss}, {Bento}, {Zhou}, {Rabus}, {Mancini}, {Henning}, {Schmidt}, {Tan},
  {Tinney}, {Wright}, {Kedziora-Chudczer}, {Bailey}, {Suc}, {Durkan},
  {L{\'a}z{\'a}r}, {Papp}, \& {S{\'a}ri}}]{deVal-Borro:2016}
{de Val-Borro}, M., {Bakos}, G.~{\'A}., {Brahm}, R., {et~al.} 2016, The
  Astronomical Journal, 152, 161, \dodoi{10.3847/0004-6256/152/6/161}

\bibitem[{{Deleuil} {et~al.}(2014){Deleuil}, {Almenara}, {Santerne}, {Barros},
  {Havel}, {H{\'e}brard}, {Bonomo}, {Bouchy}, {Bruno}, {Damiani}, {D{\'\i}az},
  {Montagnier}, \& {Moutou}}]{Deleuil:2014}
{Deleuil}, M., {Almenara}, J.~M., {Santerne}, A., {et~al.} 2014, Astronomy and
  Astrophysics, 564, A56, \dodoi{10.1051/0004-6361/201323017}

\bibitem[{{Delrez} {et~al.}(2016){Delrez}, {Santerne}, {Almenara}, {Anderson},
  {Collier-Cameron}, {D{\'\i}az}, {Gillon}, {Hellier}, {Jehin}, {Lendl},
  {Maxted}, {Neveu-VanMalle}, {Pepe}, {Pollacco}, {Queloz}, {S{\'e}gransan},
  {Smalley}, {Smith}, {Triaud}, {Udry}, {Van Grootel}, \& {West}}]{Delrez:2016}
{Delrez}, L., {Santerne}, A., {Almenara}, J.~M., {et~al.} 2016, Monthly Notices
  of the Royal Astronomical Society, 458, 4025, \dodoi{10.1093/mnras/stw522}

\bibitem[{{Duguid} {et~al.}(2020){Duguid}, {Barker}, \& {Jones}}]{Duguid:20}
{Duguid}, C.~D., {Barker}, A.~J., \& {Jones}, C.~A. 2020, \mnras, 497, 3400,
  \dodoi{10.1093/mnras/staa2216}

\bibitem[{{Essick} \& {Weinberg}(2016)}]{Essick:2016}
{Essick}, R., \& {Weinberg}, N.~N. 2016, The Astrophysical Journal, 816, 18,
  \dodoi{10.3847/0004-637X/816/1/18}

\bibitem[{{Esteves} {et~al.}(2015){Esteves}, {De Mooij}, \&
  {Jayawardhana}}]{Esteves:2015}
{Esteves}, L.~J., {De Mooij}, E. J.~W., \& {Jayawardhana}, R. 2015, The
  Astrophysical Journal, 804, 150, \dodoi{10.1088/0004-637X/804/2/150}

\bibitem[{{Goldreich} \& {Soter}(1966)}]{Goldreich:1966}
{Goldreich}, P., \& {Soter}, S. 1966, Icarus, 5, 375,
  \dodoi{10.1016/0019-1035(66)90051-0}

\bibitem[{Goodman \& Dickson(1998)}]{Goodman:1998}
Goodman, J., \& Dickson, E.~S. 1998, The Astrophysical Journal, 507, 938.
\newblock \url{http://stacks.iop.org/0004-637X/507/i=2/a=938}

\bibitem[{{Guo} {et~al.}(2023){Guo}, {Ogilvie}, \& {Barker}}]{Guo:2023}
{Guo}, Z., {Ogilvie}, G.~I., \& {Barker}, A.~J. 2023, Monthly Notices of the
  Royal Astronomical Society, 521, 1353, \dodoi{10.1093/mnras/stad569}

\bibitem[{{Hamer} \& {Schlaufman}(2019)}]{Hamer:2019}
{Hamer}, J.~H., \& {Schlaufman}, K.~C. 2019, The Astronomical Journal, 158,
  190, \dodoi{10.3847/1538-3881/ab3c56}

\bibitem[{{Harre} {et~al.}(2023){Harre}, {Smith}, {Barros}, {Bou{\'e}},
  {Csizmadia}, {Ehrenreich}, {Flor{\'e}n}, {Fortier}, {Maxted}, {Hooton},
  {Akinsanmi}, {Serrano}, {Ros{\'a}rio}, {Demory}, {Jones}, {Laskar},
  {Adibekyan}, {Alibert}, {Alonso}, {Anderson}, {Anglada}, {Asquier},
  {B{\'a}rczy}, {Barrado y Navascues}, {Baumjohann}, {Beck}, {Beck}, {Benz},
  {Billot}, {Biondi}, {Bonfanti}, {Bonfils}, {Brandeker}, {Broeg}, {Cabrera},
  {Cessa}, {Charnoz}, {Collier Cameron}, {Davies}, {Deleuil}, {Delrez},
  {Demangeon}, {Erikson}, {Fossati}, {Fridlund}, {Gandolfi}, {Gillon},
  {G{\"u}del}, {Hellier}, {Heng}, {Hoyer}, {Isaak}, {Kiss}, {Lecavelier des
  Etangs}, {Lendl}, {Lovis}, {Luntzer}, {Magrin}, {Nascimbeni}, {Olofsson},
  {Ottensamer}, {Pagano}, {Pall{\'e}}, {Persson}, {Peter}, {Piotto},
  {Pollacco}, {Queloz}, {Ragazzoni}, {Rando}, {Rauer}, {Ribas}, {Ricker},
  {Salmon}, {Santos}, {Scandariato}, {Seager}, {S{\'e}gransan}, {Simon},
  {Sousa}, {Steller}, {Szab{\'o}}, {Thomas}, {Udry}, {Ulmer}, {Van Grootel},
  {Walton}, {Wilson}, {Winn}, \& {Wohler}}]{Harre:2023}
{Harre}, J.~V., {Smith}, A.~M.~S., {Barros}, S.~C.~C., {et~al.} 2023, Astronomy
  and Astrophysics, 669, A124, \dodoi{10.1051/0004-6361/202244529}

\bibitem[{{Hebb} {et~al.}(2009){Hebb}, {Collier-Cameron}, {Loeillet},
  {Pollacco}, {H{\'e}brard}, {Street}, {Bouchy}, {Stempels}, {Moutou},
  {Simpson}, {Udry}, {Joshi}, {West}, {Skillen}, {Wilson}, {McDonald},
  {Gibson}, {Aigrain}, {Anderson}, {Benn}, {Christian}, {Enoch}, {Haswell},
  {Hellier}, {Horne}, {Irwin}, {Lister}, {Maxted}, {Mayor}, {Norton}, {Parley},
  {Pont}, {Queloz}, {Smalley}, \& {Wheatley}}]{Hebb:2009}
{Hebb}, L., {Collier-Cameron}, A., {Loeillet}, B., {et~al.} 2009, The
  Astrophysical Journal, 693, 1920, \dodoi{10.1088/0004-637X/693/2/1920}

\bibitem[{{Hellier} {et~al.}(2019){Hellier}, {Anderson}, {Bouchy}, {Burdanov},
  {Collier Cameron}, {Delrez}, {Gillon}, {Jehin}, {Lendl}, {Nielsen}, {Maxted},
  {Pepe}, {Pollacco}, {Queloz}, {S{\'e}gransan}, {Smalley}, {Triaud}, {Udry},
  \& {West}}]{Hellier:2019}
{Hellier}, C., {Anderson}, D.~R., {Bouchy}, F., {et~al.} 2019, Monthly Notices
  of the Royal Astronomical Society, 482, 1379, \dodoi{10.1093/mnras/sty2741}

\bibitem[{{Henning} {et~al.}(2018){Henning}, {Mancini}, {Sarkis}, {Bakos},
  {Hartman}, {Bayliss}, {Bento}, {Bhatti}, {Brahm}, {Ciceri}, {Csubry}, {de
  Val-Borro}, {Espinoza}, {Fulton}, {Howard}, {Isaacson}, {Jord{\'a}n},
  {Marcy}, {Penev}, {Rabus}, {Suc}, {Tan}, {Tinney}, {Wright}, {Zhou},
  {Durkan}, {Lazar}, {Papp}, \& {Sari}}]{Henning:2018}
{Henning}, T., {Mancini}, L., {Sarkis}, P., {et~al.} 2018, The Astronomical
  Journal, 155, 79, \dodoi{10.3847/1538-3881/aaa254}

\bibitem[{{Higgins} \& {Kopal}(1968)}]{Higgins:68}
{Higgins}, T.~P., \& {Kopal}, Z. 1968, Astrophysics and Space Science, 2, 352,
  \dodoi{10.1007/BF00650913}

\bibitem[{Hindmarsh {et~al.}(2005)Hindmarsh, Brown, Grant, Lee, Serban,
  Shumaker, \& Woodward}]{Hindmarsh:05}
Hindmarsh, A.~C., Brown, P.~N., Grant, K.~E., {et~al.} 2005, ACM Transactions
  on Mathematical Software (TOMS), 31, 363

\bibitem[{{Ivanov} {et~al.}(2013){Ivanov}, {Papaloizou}, \&
  {Chernov}}]{Ivanov:2013}
{Ivanov}, P.~B., {Papaloizou}, J.~C.~B., \& {Chernov}, S.~V. 2013, Monthly
  Notices of the Royal Astronomical Society, 432, 2339,
  \dodoi{10.1093/mnras/stt595}

\bibitem[{{Ivshina} \& {Winn}(2022)}]{Ivshina:2022}
{Ivshina}, E.~S., \& {Winn}, J.~N. 2022, The Astrophysical Journal Supplement
  Series, 259, 62, \dodoi{10.3847/1538-4365/ac545b}

\bibitem[{Jackson {et~al.}(2009)Jackson, Barnes, \& Greenberg}]{Jackson:2009}
Jackson, B., Barnes, R., \& Greenberg, R. 2009, The Astrophysical Journal, 698,
  1357.
\newblock \url{http://stacks.iop.org/0004-637X/698/i=2/a=1357}

\bibitem[{Jackson {et~al.}(2008)Jackson, Greenberg, \& Barnes}]{Jackson:2008}
Jackson, B., Greenberg, R., \& Barnes, R. 2008, The Astrophysical Journal, 678,
  1396.
\newblock \url{http://stacks.iop.org/0004-637X/678/i=2/a=1396}

\bibitem[{{Jermyn} {et~al.}(2022){Jermyn}, {Bauer}, {Schwab}, {Farmer}, {Ball},
  {Bellinger}, {Dotter}, {Joyce}, {Marchant}, {Mombarg}, {Wolf}, {Wong},
  {Cinquegrana}, {Farrell}, {Smolec}, {Thoul}, {Cantiello}, {Herwig}, {Toloza},
  {Bildsten}, {Townsend}, \& {Timmes}}]{Jermyn:2022}
{Jermyn}, A.~S., {Bauer}, E.~B., {Schwab}, J., {et~al.} 2022, arXiv e-prints,
  arXiv:2208.03651, \dodoi{10.48550/arXiv.2208.03651}

\bibitem[{{Kraft}(1967)}]{Kraft:1967}
{Kraft}, R.~P. 1967, The Astrophysical Journal, 150, 551,
  \dodoi{10.1086/149359}

\bibitem[{{Lazovik}(2021)}]{Lazovik:2021}
{Lazovik}, Y.~A. 2021, Monthly Notices of the Royal Astronomical Society, 508,
  3408, \dodoi{10.1093/mnras/stab2768}

\bibitem[{{Maciejewski} {et~al.}(2016){Maciejewski}, {Dimitrov},
  {Fern{\'a}ndez}, {Sota}, {Nowak}, {Ohlert}, {Nikolov}, {Bukowiecki}, {Hinse},
  {Pall{\'e}}, {Tingley}, {Kjurkchieva}, {Lee}, \& {Lee}}]{Maciejewski:2016}
{Maciejewski}, G., {Dimitrov}, D., {Fern{\'a}ndez}, M., {et~al.} 2016,
  Astronomy and Astrophysics, 588, L6, \dodoi{10.1051/0004-6361/201628312}

\bibitem[{{Maciejewski} {et~al.}(2022){Maciejewski}, {Fern{\'a}ndez}, {Sota},
  {Amado}, {Dimitrov}, {Nikolov}, {Ohlert}, {Mugrauer}, {Bischoff}, {Heyne},
  {Hildebrandt}, {Stenglein}, {Ar{\'e}valo}, {Neira}, {Riesco}, {S{\'a}nchez
  Mart{\'\i}nez}, \& {Verdugo}}]{Maciejewski:2022}
{Maciejewski}, G., {Fern{\'a}ndez}, M., {Sota}, A., {et~al.} 2022, Astronomy
  and Astrophysics, 667, A127, \dodoi{10.1051/0004-6361/202244280}

\bibitem[{{Mannaday} {et~al.}(2022){Mannaday}, {Thakur}, {Southworth}, {Jiang},
  {Sahu}, {Mancini}, {Va{\v{n}}ko}, {Kundra}, {Gajdo{\v{s}}}, {A-thano},
  {Sariya}, {Yeh}, {Griv}, {Mkrtichian}, \& {Shlyapnikov}}]{Mannaday:2022}
{Mannaday}, V.~K., {Thakur}, P., {Southworth}, J., {et~al.} 2022, The
  Astronomical Journal, 164, 198, \dodoi{10.3847/1538-3881/ac91c2}

\bibitem[{{McCormac} {et~al.}(2020){McCormac}, {Gillen}, {Jackman}, {Brown},
  {Bayliss}, {Wheatley}, {Anderson}, {Armstrong}, {Bouchy}, {Briegal},
  {Burleigh}, {Cabrera}, {Casewell}, {Chaushev}, {Chazelas}, {Chote}, {Cooke},
  {Costes}, {Csizmadia}, {Eigm{\"u}ller}, {Erikson}, {Foxell}, {G{\"a}nsicke},
  {Goad}, {G{\"u}nther}, {Hodgkin}, {Hooton}, {Jenkins}, {Lambert}, {Lendl},
  {Longstaff}, {Louden}, {Moyano}, {Nielsen}, {Pollacco}, {Queloz}, {Rauer},
  {Raynard}, {Smith}, {Smalley}, {Soto}, {Turner}, {Udry}, {Vines}, {Walker},
  {Watson}, \& {West}}]{McCormac:2020}
{McCormac}, J., {Gillen}, E., {Jackman}, J. A.~G., {et~al.} 2020, Monthly
  Notices of the Royal Astronomical Society, 493, 126,
  \dodoi{10.1093/mnras/staa115}

\bibitem[{{McQuillan} {et~al.}(2013){McQuillan}, {Mazeh}, \&
  {Aigrain}}]{McQuillan:2013}
{McQuillan}, A., {Mazeh}, T., \& {Aigrain}, S. 2013, \apjl, 775, L11,
  \dodoi{10.1088/2041-8205/775/1/L11}

\bibitem[{{Nielsen} {et~al.}(2020){Nielsen}, {Brahm}, {Bouchy}, {Espinoza},
  {Turner}, {Rappaport}, {Pearce}, {Ricker}, {Vanderspek}, {Latham}, {Seager},
  {Winn}, {Jenkins}, {Acton}, {Bakos}, {Barclay}, {Barkaoui}, {Bhatti},
  {Brice{\~n}o}, {Bryant}, {Burleigh}, {Ciardi}, {Collins}, {Collins}, {Cooke},
  {Csubry}, {dos Santos}, {Eigm{\"u}ller}, {Fausnaugh}, {Gan}, {Gillon},
  {Goad}, {Guerrero}, {Hagelberg}, {Hart}, {Henning}, {Huang}, {Jehin},
  {Jenkins}, {Jord{\'a}n}, {Kielkopf}, {Kossakowski}, {Lavie}, {Law}, {Lendl},
  {de Leon}, {Lovis}, {Mann}, {Marmier}, {McCormac}, {Mori}, {Moyano},
  {Narita}, {Osip}, {Otegi}, {Pepe}, {Pozuelos}, {Raynard}, {Relles}, {Sarkis},
  {S{\'e}gransan}, {Seidel}, {Shporer}, {Stalport}, {Stockdale}, {Suc},
  {Tamura}, {Tan}, {Tilbrook}, {Ting}, {Trifonov}, {Udry}, {Vanderburg},
  {Wheatley}, {Wingham}, {Zhan}, \& {Ziegler}}]{Nielsen:2020}
{Nielsen}, L.~D., {Brahm}, R., {Bouchy}, F., {et~al.} 2020, Astronomy and
  Astrophysics, 639, A76, \dodoi{10.1051/0004-6361/202037941}

\bibitem[{{Oberst} {et~al.}(2017){Oberst}, {Rodriguez}, {Col{\'o}n},
  {Angerhausen}, {Bieryla}, {Ngo}, {Stevens}, {Stassun}, {Gaudi}, {Pepper},
  {Penev}, {Mawet}, {Latham}, {Heintz}, {Osei}, {Collins}, {Kielkopf},
  {Visgaitis}, {Reed}, {Escamilla}, {Yazdi}, {McLeod}, {Lunsford}, {Spencer},
  {Joner}, {Gregorio}, {Gaillard}, {Matt}, {Dumont}, {Stephens}, {Cohen},
  {Jensen}, {Calchi Novati}, {Bozza}, {Labadie-Bartz}, {Siverd}, {Lund},
  {Beatty}, {Eastman}, {Penny}, {Manner}, {Zambelli}, {Fulton}, {Stockdale},
  {DePoy}, {Marshall}, {Pogge}, {Gould}, {Trueblood}, \&
  {Trueblood}}]{Oberst:2017}
{Oberst}, T.~E., {Rodriguez}, J.~E., {Col{\'o}n}, K.~D., {et~al.} 2017, The
  Astronomical Journal, 153, 97, \dodoi{10.3847/1538-3881/153/3/97}

\bibitem[{Ogilvie(2014)}]{Ogilvie:2014}
Ogilvie, G.~I. 2014, Annual Review of Astronomy and Astrophysics, 52, 171,
  \dodoi{10.1146/annurev-astro-081913-035941}

\bibitem[{{Oliveira} {et~al.}(2019){Oliveira}, {Martioli}, \&
  {Tucci-Maia}}]{Oliveira:2019}
{Oliveira}, J.~M., {Martioli}, E., \& {Tucci-Maia}, M. 2019, Research Notes of
  the American Astronomical Society, 3, 35, \dodoi{10.3847/2515-5172/ab06c7}

\bibitem[{{Patra} {et~al.}(2017){Patra}, {Winn}, {Holman}, {Yu}, {Deming}, \&
  {Dai}}]{Patra:2017}
{Patra}, K.~C., {Winn}, J.~N., {Holman}, M.~J., {et~al.} 2017, The Astronomical
  Journal, 154, 4, \dodoi{10.3847/1538-3881/aa6d75}

\bibitem[{{Patra} {et~al.}(2020){Patra}, {Winn}, {Holman}, {Gillon},
  {Burdanov}, {Jehin}, {Delrez}, {Pozuelos}, {Barkaoui}, {Benkhaldoun},
  {Narita}, {Fukui}, {Kusakabe}, {Kawauchi}, {Terada}, {Bouma}, {Weinberg}, \&
  {Broome}}]{Patra:2020}
---. 2020, The Astronomical Journal, 159, 150, \dodoi{10.3847/1538-3881/ab7374}

\bibitem[{{Paxton} {et~al.}(2011){Paxton}, {Bildsten}, {Dotter}, {Herwig},
  {Lesaffre}, \& {Timmes}}]{Paxton:11}
{Paxton}, B., {Bildsten}, L., {Dotter}, A., {et~al.} 2011, \apjs, 192, 3,
  \dodoi{10.1088/0067-0049/192/1/3}

\bibitem[{{Paxton} {et~al.}(2013){Paxton}, {Cantiello}, {Arras}, {Bildsten},
  {Brown}, {Dotter}, {Mankovich}, {Montgomery}, {Stello}, {Timmes}, \&
  {Townsend}}]{Paxton:13}
{Paxton}, B., {Cantiello}, M., {Arras}, P., {et~al.} 2013, \apjs, 208, 4,
  \dodoi{10.1088/0067-0049/208/1/4}

\bibitem[{{Paxton} {et~al.}(2015){Paxton}, {Marchant}, {Schwab}, {Bauer},
  {Bildsten}, {Cantiello}, {Dessart}, {Farmer}, {Hu}, {Langer}, {Townsend},
  {Townsley}, \& {Timmes}}]{Paxton:15}
{Paxton}, B., {Marchant}, P., {Schwab}, J., {et~al.} 2015, \apjs, 220, 15,
  \dodoi{10.1088/0067-0049/220/1/15}

\bibitem[{{Paxton} {et~al.}(2018){Paxton}, {Schwab}, {Bauer}, {Bildsten},
  {Blinnikov}, {Duffell}, {Farmer}, {Goldberg}, {Marchant}, {Sorokina},
  {Thoul}, {Townsend}, \& {Timmes}}]{Paxton:18}
{Paxton}, B., {Schwab}, J., {Bauer}, E.~B., {et~al.} 2018, \apjs, 234, 34,
  \dodoi{10.3847/1538-4365/aaa5a8}

\bibitem[{{Paxton} {et~al.}(2019){Paxton}, {Smolec}, {Schwab}, {Gautschy},
  {Bildsten}, {Cantiello}, {Dotter}, {Farmer}, {Goldberg}, {Jermyn}, {Kanbur},
  {Marchant}, {Thoul}, {Townsend}, {Wolf}, {Zhang}, \& {Timmes}}]{Paxton:19}
{Paxton}, B., {Smolec}, R., {Schwab}, J., {et~al.} 2019, \apjs, 243, 10,
  \dodoi{10.3847/1538-4365/ab2241}

\bibitem[{{Penev} {et~al.}(2018){Penev}, {Bouma}, {Winn}, \&
  {Hartman}}]{Penev:2018}
{Penev}, K., {Bouma}, L.~G., {Winn}, J.~N., \& {Hartman}, J.~D. 2018, The
  Astronomical Journal, 155, 165, \dodoi{10.3847/1538-3881/aaaf71}

\bibitem[{{Penev} {et~al.}(2016){Penev}, {Hartman}, {Bakos}, {Ciceri}, {Brahm},
  {Bayliss}, {Bento}, {Jord{\'a}n}, {Csubry}, {Bhatti}, {de Val-Borro},
  {Espinoza}, {Zhou}, {Mancini}, {Rabus}, {Suc}, {Henning}, {Schmidt}, {Noyes},
  {L{\'a}z{\'a}r}, {Papp}, \& {S{\'a}ri}}]{Penev:2016}
{Penev}, K., {Hartman}, J.~D., {Bakos}, G.~{\'A}., {et~al.} 2016, The
  Astronomical Journal, 152, 127, \dodoi{10.3847/0004-6256/152/5/127}

\bibitem[{{Ros{\'a}rio} {et~al.}(2022){Ros{\'a}rio}, {Barros}, {Demangeon}, \&
  {Santos}}]{Rosario:2022}
{Ros{\'a}rio}, N.~M., {Barros}, S.~C.~C., {Demangeon}, O.~D.~S., \& {Santos},
  N.~C. 2022, Astronomy and Astrophysics, 668, A114,
  \dodoi{10.1051/0004-6361/202244513}

\bibitem[{Schenk {et~al.}(2001)Schenk, Arras, Flanagan, Teukolsky, \&
  Wasserman}]{Schenk:2001}
Schenk, A.~K., Arras, P., Flanagan, E.~E., Teukolsky, S.~A., \& Wasserman, I.
  2001, Phys. Rev. D, 65, 024001, \dodoi{10.1103/PhysRevD.65.024001}

\bibitem[{{Schenk} {et~al.}(2001){Schenk}, {Arras}, {Flanagan}, {Teukolsky}, \&
  {Wasserman}}]{Schenk:02}
{Schenk}, A.~K., {Arras}, P., {Flanagan}, {\'E}.~{\'E}., {Teukolsky}, S.~A., \&
  {Wasserman}, I. 2001, \prd, 65, 024001, \dodoi{10.1103/PhysRevD.65.024001}

\bibitem[{{Southworth}(2012)}]{Southworth:2012}
{Southworth}, J. 2012, Monthly Notices of the Royal Astronomical Society, 426,
  1291, \dodoi{10.1111/j.1365-2966.2012.21756.x}

\bibitem[{{Spake} {et~al.}(2016){Spake}, {Brown}, {Doyle}, {H{\'e}brard},
  {McCormac}, {Armstrong}, {Pollacco}, {G{\'o}mez Maqueo Chew}, {Anderson},
  {Barros}, {Bouchy}, {Boumis}, {Bruno}, {Collier Cameron}, {Courcol},
  {Davies}, {Faedi}, {Hellier}, {Kirk}, {Lam}, {Liakos}, {Louden}, {Maxted},
  {Osborn}, {Palle}, {Prieto Arranz}, {Udry}, {Walker}, {West}, \&
  {Wheatley}}]{Spake:2016}
{Spake}, J.~J., {Brown}, D.~J.~A., {Doyle}, A.~P., {et~al.} 2016, Publications
  of the Astronomical Society of the Pacific, 128, 024401,
  \dodoi{10.1088/1538-3873/128/960/024401}

\bibitem[{{Stassun} {et~al.}(2017){Stassun}, {Collins}, \&
  {Gaudi}}]{Stassun:2017}
{Stassun}, K.~G., {Collins}, K.~A., \& {Gaudi}, B.~S. 2017, The Astronomical
  Journal, 153, 136, \dodoi{10.3847/1538-3881/aa5df3}

\bibitem[{{Storch} \& {Lai}(2014)}]{Storch:2014}
{Storch}, N.~I., \& {Lai}, D. 2014, mnras, 438, 1526,
  \dodoi{10.1093/mnras/stt2292}

\bibitem[{{Sun} {et~al.}(2018){Sun}, {Arras}, {Weinberg}, {Troup}, \&
  {Majewski}}]{Sun:2018}
{Sun}, M., {Arras}, P., {Weinberg}, N.~N., {Troup}, N.~W., \& {Majewski}, S.~R.
  2018, Monthly Notices of the Royal Astronomical Society, 481, 4077,
  \dodoi{10.1093/mnras/sty2464}

\bibitem[{{Teitler} \& {K{\"o}nigl}(2014)}]{Teitler:2014}
{Teitler}, S., \& {K{\"o}nigl}, A. 2014, \apj, 786, 139,
  \dodoi{10.1088/0004-637X/786/2/139}

\bibitem[{{Terquem} {et~al.}(1998){Terquem}, {Papaloizou}, {Nelson}, \&
  {Lin}}]{Terquem:1998}
{Terquem}, C., {Papaloizou}, J.~C.~B., {Nelson}, R.~P., \& {Lin}, D.~N.~C.
  1998, \apj, 502, 788, \dodoi{10.1086/305927}

\bibitem[{Townsend {et~al.}(2018)Townsend, Goldstein, \& Zweibel}]{Townsend:18}
Townsend, R. H.~D., Goldstein, J., \& Zweibel, E.~G. 2018, Monthly Notices of
  the Royal Astronomical Society, 475, 879, \dodoi{10.1093/mnras/stx3142}

\bibitem[{{Townsend} \& {Teitler}(2013)}]{Townsend:13}
{Townsend}, R.~H.~D., \& {Teitler}, S.~A. 2013, \mnras, 435, 3406,
  \dodoi{10.1093/mnras/stt1533}

\bibitem[{{Turner} {et~al.}(2016){Turner}, {Anderson}, {Collier Cameron},
  {Delrez}, {Evans}, {Gillon}, {Hellier}, {Jehin}, {Lendl}, {Maxted}, {Pepe},
  {Pollacco}, {Queloz}, {S{\'e}gransan}, {Smalley}, {Smith}, {Triaud}, {Udry},
  \& {West}}]{Turner:2016}
{Turner}, O.~D., {Anderson}, D.~R., {Collier Cameron}, A., {et~al.} 2016,
  Publications of the Astronomical Society of the Pacific, 128, 064401,
  \dodoi{10.1088/1538-3873/128/964/064401}

\bibitem[{{Van Hoolst}(1994)}]{VanHoolst:1994}
{Van Hoolst}, T. 1994, aap, 286, 879

\bibitem[{{Vines} {et~al.}(2019){Vines}, {Jenkins}, {Acton}, {Briegal},
  {Bayliss}, {Bouchy}, {Belardi}, {Bryant}, {Burleigh}, {Cabrera}, {Casewell},
  {Chaushev}, {Cooke}, {Csizmadia}, {Eigm{\"u}ller}, {Erikson}, {Foxell},
  {Gill}, {Gillen}, {Goad}, {Jackman}, {King}, {Louden}, {McCormac}, {Moyano},
  {Nielsen}, {Pollacco}, {Queloz}, {Rauer}, {Raynard}, {Smith}, {Soto},
  {Tilbrook}, {Titz-Weider}, {Turner}, {Udry}, {Walker}, {Watson}, {West}, \&
  {Wheatley}}]{Vines:2019}
{Vines}, J.~I., {Jenkins}, J.~S., {Acton}, J.~S., {et~al.} 2019, Monthly
  Notices of the Royal Astronomical Society, 489, 4125,
  \dodoi{10.1093/mnras/stz2349}

\bibitem[{{Vissapragada} {et~al.}(2022){Vissapragada}, {Chontos},
  {Greklek-McKeon}, {Knutson}, {Dai}, {P{\'e}rez Gonz{\'a}lez}, {Grunblatt},
  {Huber}, \& {Saunders}}]{Vissapragada:2022}
{Vissapragada}, S., {Chontos}, A., {Greklek-McKeon}, M., {et~al.} 2022, The
  Astrophysical Journal, 941, L31, \dodoi{10.3847/2041-8213/aca47e}

\bibitem[{{Weinberg} {et~al.}(2012){Weinberg}, {Arras}, {Quataert}, \&
  {Burkart}}]{Weinberg:12}
{Weinberg}, N.~N., {Arras}, P., {Quataert}, E., \& {Burkart}, J. 2012, \apj,
  751, 136, \dodoi{10.1088/0004-637X/751/2/136}

\bibitem[{Weinberg {et~al.}(2012)Weinberg, Arras, Quataert, \&
  Burkart}]{Weinberg:2012}
Weinberg, N.~N., Arras, P., Quataert, E., \& Burkart, J. 2012, The
  Astrophysical Journal, 751, 136.
\newblock \url{http://stacks.iop.org/0004-637X/751/i=2/a=136}

\bibitem[{{Weinberg} {et~al.}(2017){Weinberg}, {Sun}, {Arras}, \&
  {Essick}}]{Weinberg:2017}
{Weinberg}, N.~N., {Sun}, M., {Arras}, P., \& {Essick}, R. 2017, The
  Astrophysical Journal, 849, L11, \dodoi{10.3847/2041-8213/aa9113}

\bibitem[{{West} {et~al.}(2016){West}, {Hellier}, {Almenara}, {Anderson},
  {Barros}, {Bouchy}, {Brown}, {Collier Cameron}, {Deleuil}, {Delrez}, {Doyle},
  {Faedi}, {Fumel}, {Gillon}, {G{\'o}mez Maqueo Chew}, {H{\'e}brard}, {Jehin},
  {Lendl}, {Maxted}, {Pepe}, {Pollacco}, {Queloz}, {S{\'e}gransan}, {Smalley},
  {Smith}, {Southworth}, {Triaud}, \& {Udry}}]{West:2016}
{West}, R.~G., {Hellier}, C., {Almenara}, J.~M., {et~al.} 2016, Astronomy and
  Astrophysics, 585, A126, \dodoi{10.1051/0004-6361/201527276}

\bibitem[{{Wong} {et~al.}(2021){Wong}, {Shporer}, {Zhou}, {Kitzmann},
  {Komacek}, {Tan}, {Tronsgaard}, {Buchhave}, {Vissapragada}, {Greklek-McKeon},
  {Rodriguez}, {Ahlers}, {Quinn}, {Furlan}, {Howell}, {Bieryla}, {Heng},
  {Knutson}, {Collins}, {McLeod}, {Berlind}, {Brown}, {Calkins}, {de Leon},
  {Esparza-Borges}, {Esquerdo}, {Fukui}, {Gan}, {Girardin}, {Gnilka}, {Ikoma},
  {Jensen}, {Kielkopf}, {Kodama}, {Kurita}, {Lester}, {Lewin}, {Marino},
  {Murgas}, {Narita}, {Pall{\'e}}, {Schwarz}, {Stassun}, {Tamura}, {Watanabe},
  {Benneke}, {Ricker}, {Latham}, {Vanderspek}, {Seager}, {Winn}, {Jenkins},
  {Caldwell}, {Fong}, {Huang}, {Mireles}, {Schlieder}, {Shiao}, \& {Noel
  Villase{\~n}or}}]{Wong:2021}
{Wong}, I., {Shporer}, A., {Zhou}, G., {et~al.} 2021, The Astronomical Journal,
  162, 256, \dodoi{10.3847/1538-3881/ac26bd}

\bibitem[{{Wu} \& {Goldreich}(2001)}]{Wu:2001}
{Wu}, Y., \& {Goldreich}, P. 2001, \apj, 546, 469, \dodoi{10.1086/318234}

\bibitem[{{Yee} {et~al.}(2020){Yee}, {Winn}, {Knutson}, {Patra},
  {Vissapragada}, {Zhang}, {Holman}, {Shporer}, \& {Wright}}]{Yee:2020}
{Yee}, S.~W., {Winn}, J.~N., {Knutson}, H.~A., {et~al.} 2020, The Astrophysical
  Journal, 888, L5, \dodoi{10.3847/2041-8213/ab5c16}

\bibitem[{{Yee} {et~al.}(2023){Yee}, {Winn}, {Hartman}, {Bouma}, {Zhou},
  {Quinn}, {Latham}, {Bieryla}, {Rodriguez}, {Collins}, {Alfaro}, {Barkaoui},
  {Beard}, {Belinski}, {Benkhaldoun}, {Benni}, {Bernacki}, {Boyle}, {Butler},
  {Caldwell}, {Chontos}, {Christiansen}, {Ciardi}, {Collins}, {Conti}, {Crane},
  {Daylan}, {Dressing}, {Eastman}, {Essack}, {Evans}, {Everett},
  {Fajardo-Acosta}, {For{\'e}s-Toribio}, {Furlan}, {Ghachoui}, {Gillon},
  {Hellier}, {Helm}, {Howard}, {Howell}, {Isaacson}, {Jehin}, {Jenkins},
  {Jensen}, {Kielkopf}, {Laloum}, {Leonhardes-Barboza}, {Lewin}, {Logsdon},
  {Lubin}, {Lund}, {MacDougall}, {Mann}, {Maslennikova}, {Massey}, {McLeod},
  {Mu{\~n}oz}, {Newman}, {Orlov}, {Plavchan}, {Popowicz}, {Pozuelos},
  {Pritchard}, {Radford}, {Reefe}, {Ricker}, {Rudat}, {Safonov}, {Schwarz},
  {Schweiker}, {Scott}, {Seager}, {Shectman}, {Stockdale}, {Tan}, {Teske},
  {Thomas}, {Timmermans}, {Vanderspek}, {Vermilion}, {Watanabe}, {Weiss},
  {West}, {Van Zandt}, {Zejmo}, \& {Ziegler}}]{Yee:2023}
{Yee}, S.~W., {Winn}, J.~N., {Hartman}, J.~D., {et~al.} 2023, The Astrophysical
  Journal Supplement Series, 265, 1, \dodoi{10.3847/1538-4365/aca286}

\bibitem[{{Yu} {et~al.}(2020){Yu}, {Weinberg}, \& {Fuller}}]{Yu:2020}
{Yu}, H., {Weinberg}, N.~N., \& {Fuller}, J. 2020, Monthly Notices of the Royal
  Astronomical Society, 496, 5482, \dodoi{10.1093/mnras/staa1858}

\end{thebibliography}
\bibliographystyle{aasjournal}

\end{document}